\documentclass[11pt]{article}
\usepackage{jheppub}
\usepackage{amsmath,amssymb,physics,bm,mathtools}
\usepackage{microtype}
\usepackage{appendix}
\usepackage{color}

\newcommand{\AdS}{\mathrm{AdS}}

\newcommand{\Lie}{\mathcal{L}}

\newcommand{\fm}{\mathfrak{m}}
\newcommand{\so}{\mathfrak{so}}
\newcommand{\iso}{\mathfrak{iso}}
\newcommand{\g}{\mathfrak{g}}
\newcommand{\h}{\mathfrak{h}}

\newcommand{\GL}{\operatorname{GL}}
\newcommand{\SL}{\operatorname{SL}}
\newcommand{\SO}{\operatorname{SO}}
\newcommand{\Ad}{\operatorname{Ad}}
\newcommand{\ad}{\operatorname{ad}}
\newcommand{\RR}{\mathbb{R}}
\newcommand{\eO}{\mathcal{O}}
\newcommand{\vol}{\operatorname{vol}}
\newcommand\ext[1]{\scalebox{.8}[1]{$\bigwedge^{\!#1}$}}
\newcommand{\nablag}{\nabla^{(g)}}

\usepackage{tikz,pgfplots}
\usepgfplotslibrary{fillbetween}
\usetikzlibrary{calc,arrows,arrows.meta,cd,shapes.geometric,positioning,through,decorations.markings,decorations.text}
\tikzset{>=latex}
\title{Massive and massless particles in Mielke--Baekler geometries}

\author[a]{Carles Batlle,}
\author[b]{Roberto Casalbuoni,}
\author[b]{Daniele Dominici,}
\author[c]{José Figueroa-O'Farrill,}
\author[d]{and Joaquim Gomis}

\affiliation[a]{Institut d'Organització i Control i Departament de
  Matemàtiques, Universitat Politècnica de Catalunya, EPSEVG,
  Av. V. Balaguer 1,  Vilanova i la Geltrú, 08800 Spain}
\affiliation[b]{Department of Physics and Astronomy, University of Florence and INFN, 50019 Sesto Fiorentino, Italy}
\affiliation[c]{Maxwell Institute and School of Mathematics, The
  University of Edinburgh, James Clerk Maxwell Building, Peter Guthrie
  Tait Road, King's Buildings, Edinburgh EH9 3FD, United Kingdom}
\affiliation[d]{Emeritus Professor, Universitat de Barcelona, Gran Via
  de les Corts Catalanes 585, 08007 Barcelona, Spain}
\emailAdd{carles.batlle@upc.edu}
\emailAdd{casalbuoni@fi.infn.it}
\emailAdd{dominici@fi.infn.it}
\emailAdd{j.m.figueroa@ed.ac.uk}
\emailAdd{joaquim.gomis@ub.edu}

\abstract{The Mielke--Baekler geometries are three-dimensional
  reductive homogeneous spacetimes together with a choice of invariant
  connection which is compatible with a lorentzian metric.  The
  spacetimes generalise Minkowski and (anti)de~Sitter spacetimes in
  that the invariant metric connection can have torsion, a peculiarity
  of three dimensions.  Using coadjoint orbits and the techniques of
  nonlinear realisations, we construct worldline actions for massive
  and massless spinning particles moving in these spacetimes.  We pay
  particular attention to the so-called teleparallel branch, in which
  the curvature of the invariant connection vanishes.   Apart from the trivial Minkowski case, this singles
  out anti-de~Sitter spacetime, as the only of these lorentzian
  manifolds admitting an invariant Weitzenböck connection; that is, a
  flat connection with torsion.  The introduction of a Wess--Zumino
  term describing spin has, as a main consequence, the appearance of
  dynamical sectors (denoted ``regular'' and ``critical''), with a
  different number of physical degrees of freedom. In particular, in
  the massive case, we discuss the formulation of the dynamics in
  terms of either the Weitzenböck or the Levi-Civita connection, and
  the emergence of a Papapetrou-type forcing term in the critical sector. 
   For the massless particle we study the Noether
  symmetries of the action, which, for the spinless case, include the
  conformal transformations. For nonzero spin, only the Killing subset
  survives as genuine Noether transformations in the regular sector,
  while in the critical sector any conformal Killing contribution can
  be set to zero by a gauge transformation.}
  

\begin{document}
\maketitle
\flushbottom
	
\section{Introduction}
\label{sec:intro}

Three-dimensional gravity provides an unusually clean arena for
studying the relation between spacetime geometry and matter. In $2+1$
dimensions pure gravity has no local propagating degrees of freedom,
and many theories admit a reformulation as gauge theories; the
canonical example being the Chern--Simons (CS) description of $2+1$
Einstein gravity \cite{Achucarro:1986vz,Witten:1988hc}.

A natural extension of $2+1$ gravity is to allow for torsion. In a
Cartan formulation one treats a coframe $\theta^A$ and spin connection
$\omega^A$ as independent variables. The Mielke--Baekler (MB) model
\cite{Mielke:1991nn, Blagojevic:2002, BlagojevicCvetkovic:2004,
  Giacomini:2006dr,Cvetkovic:2007sr,Barriga:2026awj} is a well-studied class of such theories and
admits as solutions homogeneous spacetimes with invariant connections
with both curvature and torsion.  In \cite{Geiller:2020edh} it was
shown that MB gravity is associated with a deformation of the Poincaré
algebra in which the translation sector acquires a curvature term and
a torsional term, which is what is usually called the MB algebra.
These Lie algebras are isomorphic to the Lie algebras of
isometries of (anti) de Sitter and Minkowski spacetimes, but they
suggest a description of these spacetimes in which the canonical
connection is different from the Levi-Civita connection.

Torsion is expected to couple naturally to spinning matter
\cite{Hehl:1976kj,Shapiro:2002}.  A particularly simple probe is a
point particle. In this paper we derive particle actions propagating
on these MB geometries using the method of coadjoint orbits, nonlinear
realisations and coset constructions
\cite{Coleman:1969sm, Callan:1969sn, Volkov:1973vd, Gomis:2006xw,
  Bergshoeff:2022eog}. This method systematically produces invariant
worldline lagrangians from the Maurer--Cartan (MC) forms and
automatically yields the effective background geometry.

A special case of MB geometry are the teleparallel geometries, where
the curvature vanishes. Teleparallel geometry replaces curvature by
torsion using a Weitzenböck connection built from the coframe
\cite{Aldrovandi:2013wha, Pereira:2019}. In such a geometry, we show
explicitly how the spinning particle dynamics can be written as an
autoparallel equation for the Weitzenböck connection. In the presence
of spin, the dynamics have two sectors, which we call ``regular'' and
``critical'' with a different number of physical degrees of freedom.
For particle models in AdS$_3$ the presence of sectors was noticed in
\cite{Batlle:2014sca,Batlle:2025byv}. In the massive case, we show
that in the critical sector, the dynamics admit an alternative
description in terms of the Levi-Civita connection but involving a
spin-curvature forcing term à la Papapetrou \cite{Mathisson:1937,
  Papapetrou:1951, Tulczyjew:1959}. In the regular sector that
spin-curvature term turns out to vanish. We also study massless
particles and determine the Noether symmetries of the action, which in
the spinless case, include conformal transformations. In the presence
of spin, the conformal transformations are not symmetries of the
regular sector, whereas they can be gauged away in the critical
sector.

We now give a detailed summary of the contents of the paper.

\paragraph{Outline of the paper.}

We start in Section~\ref{sec:MBalg} by motivating the Mielke--Baekler
family of Lie algebras and discussing some of their properties and
those of the corresponding homogeneous spacetimes.  We start in
Section~\ref{sec:maxim-symm-lorentz} by reviewing the invariant
connections of the three-dimensional lorentzian space forms: (anti) de
Sitter and Minkowski spacetimes.  This is only interesting in three
dimensions, since for any other dimension the only invariant
connection is the Levi-Civita connection.  In
Section~\ref{sec:MB-algebras} we show that given an invariant
connection $\nabla$ (determined uniquely by its invariant torsion and
curvature), there is a choice of reductive split in the Lie algebra of
isometries whose corresponding canonical connection agrees with
$\nabla$.  These are the so-called Mielke--Baekler (MB) algebras.  (In
Appendix~\ref{sec:deform-lie-algebr} we show how these algebras can be
re-interpreted in terms of Lie algebra deformations.)  In
Sections~\ref{sec:central-extensions} and \ref{sec:quadratic-casimirs}
we study the central extensions and some of the Casimirs of the MB
algebras.  In Section~\ref{sec:flat-inv-conns} we ask whether there
are any invariant Weitzenböck (i.e., flat) connections and recover the
well-known answers that Minkowski spacetime has precisely one such
connection (the Levi-Civita connection) and anti~de~Sitter spacetime
has two such connections with opposite torsions, whose existence is
explained by the fact that $\AdS_3$ is locally isometric to
$\SL(2,\RR)$ with a bi-invariant metric.  In
Section~\ref{sec:homog-geom} we study the homogeneous geometries
associated to the MB algebras and in particular exhibit manifestly
conformally-flat coordinates which will be useful in the description
of massless particles.

In Section~\ref{sec:dyn-MB} we study the dynamics of massive particles
in the MB geometries.  In Section~\ref{sec:particles-from-orbits} we
briefly review how to obtain particle dynamics from coadjoint orbits.
In Section~\ref{sec:massive-particles} we focus on coadjoint orbits
corresponding to massive, spinning particles and show that there are
two kinds of coadjoint orbits: a four-dimensional generic family of
orbits and a locus consisting of two-dimensional orbits.  This
dichotomy manifests itself in the existence of two sectors with
different dynamics: the ``regular'' sector (corresponding to the
generic orbits) and the ``critical'' sector (corresponding to the
smaller orbits).  From Section~\ref{sec:massive-teleparallel} onwards
we focus solely on the so-called ``teleparallel'' case, where the
invariant connection is flat but with non-zero torsion. (In
Appendix~\ref{sec:teleparallelreview} we include a brief review of
teleparallel geometry.)  The teleparallel condition singles out the
$\AdS_3$ geometry: the flat connection in Minkowski spacetime is the
Levi-Civita connection, which has no torsion and although de Sitter
spacetime does admit flat connections, none of them are invariant.  We
determine the lagrangian for a particle with mass $m$ and spin $s$: the
case of zero spin being treated in Section~\ref{sec:massive-spinless},
while the case of nonzero spin is treated in
Section~\ref{sec:massive-spinning}.  In both cases, the particle momentum is parallel transported by 
 the (transposed) 
Weitzenböck connection.  In the case of nonzero spin and in the
critical sector, there is an alternative description of the physical
trajectories as those satisfying a Papapetrou-like equation: a
modification of the Levi-Civita geodesic equation by a spin-curvature
forcing term.  Finally, in Section~\ref{sec:massive-canonical} we
study the canonical structure of the massive spinning particle in the
teleparallel case.  In particular we see how the nature of the
constraints depends on whether we are in the regular or critical
sectors.  The analysis of the reduced space again recovers the
dimension count of the coadjoint orbits.  Some of the details of the
calculations in this section have been relegated to
Appendix~\ref{sec:gf}.

In Section~\ref{sec:massless-particles} we study the dynamics of
massless spinning particles in the teleparallel case. In
Section~\ref{sec:massl-part-orbits} we determine the relevant
coadjoint orbits and again notice that there are two kinds of
coadjoint orbits, as in the massive case, each one giving rise to a
different dynamical sector. In Section~\ref{sect3} we give some
formulae for the homogeneous geometry of the teleparallel MB spacetime
in a convenient basis for the study of the massless dynamics. The
corresponding lagrangians are determined in Section~\ref{sec:massless}
and the canonical analysis (including the determination of the
constraints in both regular and critical sectors) is performed in
Section~\ref{sec:canonical-analysis}. In
Section~\ref{sec:conformal-symmetry} we determine a class of Noether
symmetries of the massless MB particle which includes, only in the
case of zero spin, the conformal symmetries of the spacetime. This
uses an identity which is proved in
Appendix~\ref{sec:useful-identity}. Finally in
Section~\ref{sec:algebra} we explicitly work out the conformal MB
algebra of the spinless massless particle.

The paper concludes in Section~\ref{sec:concl} where we summarise the
main results and mention some open problems suggested by them.

\section{A family of Lie algebras and their associated geometries}
\label{sec:MBalg}

In this section we will motivate a family of Lie algebras which will
be central to this paper and discuss the properties of their
associated homogeneous spacetimes.

\subsection{Maximally symmetric lorentzian spacetimes and their invariant connections}
\label{sec:maxim-symm-lorentz}

We depart from the well-known fact that there are (up to local
isometry and up to homothety) three distinct spatially isotropic
lorentzian manifolds with maximal symmetry, distinguished by the sign
of the (constant) Ricci scalar of the invariant metric: de Sitter
spacetime (positive), anti de Sitter spacetime (negative) and
Minkowski spacetime (zero).  They are all three lorentzian symmetric
spaces and they can be described infinitesimally by their Klein pairs
$(\g,\h)$, where in all cases $\h \cong \so(n-1,1)$ and $\g =
\so(n,1)$ for de Sitter, $\g = \so(n-1,2)$ for anti de Sitter and $\g
= \iso(n-1,1)$ for Minkowski, where $n$ is the dimension of the
manifold.

Being symmetric spaces, the Klein pair $(\g,\h)$ is, in particular,
reductive; that is, there exists a complement $\fm$ to $\h$ in $\g$,
which is stable under the adjoint action of $\h$: $\g = \h \oplus \fm$
and $[\h,\fm] \subset \fm$.  In addition, because the split is
symmetric, $[\fm,\fm] \subset \h$.  Since the metric is invariant, so
is the Levi-Civita connection and it is a natural question whether there are
other invariant connections. Invariant connections on a reductive
homogeneous space are characterised \cite{MR0059050} by their Nomizu
maps: $\h$-equivariant bilinear maps $N \colon \fm \times \fm \to
\fm$.  The torsion and curvature tensors of the invariant connection
associated with  a Nomizu map $N$ are also invariant and given at the
origin by the expressions
\begin{equation}
  \label{eq:Nomizu-torsion}
  T(X,Y) = N(X,Y) - N(Y,X) - [X,Y]_\fm
\end{equation}
and
\begin{equation}
  \label{eq:Nomizu-curvature}
  R(X,Y)Z = N(X,N(Y,Z)) - N(Y,N(X,Z)) - N([X,Y]_\fm, Z) - [[X,Y]_\h,Z],
\end{equation}
for all $X,Y,Z \in \fm$.  Every reductive homogeneous space has two
canonical connections (called of the first and second kind in Nomizu's
original paper \cite{MR0059050}): the one with vanishing torsion and
the one with zero Nomizu map.  It is customary nowadays to call the
invariant connection with zero Nomizu map \emph{the} canonical
connection.  The torsion and curvature tensors for the canonical
connection take a particularly simple form:
\begin{equation}
  \label{eq:Torsion-Curvature-Canonical}
  T(X,Y) = - [X,Y]_\fm \qquad\text{and}\qquad R(X,Y)Z = - [[X,Y]_\h, Z].
\end{equation}
If the reductive split $\g = \h \oplus \fm$ is also symmetric, so that
$[\fm,\fm] \subset \h$, then the canonical connection has no torsion
and hence, in the (pseudo)riemannian setting, agrees with the
Levi-Civita connection.

Spatially isotropic homogeneous kinematical spacetimes were classified
in \cite{Figueroa-OFarrill:2018ilb} and their invariant connections
were classified in \cite{Figueroa-OFarrill:2019sex}.  In particular,
in \cite[§4.1]{Figueroa-OFarrill:2019sex} the Nomizu maps for the
lorentzian spacetimes above were classified.  In all dimensions
$n\neq 3$, the only invariant connection is the Levi-Civita
connection, but in dimension $n=3$ there is a one-parameter family.
This is intimately related to the fact that only in dimension $n=3$,
the adjoint and vector representations of $\so(n-1,1)$ are equivalent.

Let us set $n=3$ from now on.  The Lie algebras $\g$ in the Klein pair
are spanned by $J_{AB}$ and $P_A$, where $A,B \in \{0,1,2\}$, with Lie
brackets
\begin{equation}
  \label{eq:Lie-algebra}
  \begin{split}
    [J_{AB},J_{CD}] &= \eta_{BC} J_{AD}- \eta_{AC} J_{BD} - \eta_{BD} J_{AC} + \eta_{AD} J_{BC}\\
    [J_{AB},P_C] &= \eta_{BC} P_A - \eta_{AC} P_B\\
    [P_A, P_B] &= -\Lambda J_{AB},
  \end{split}
\end{equation}
where the inner product $\eta_{AB} = \operatorname{diag}(-1,1,1)$ and
where we have introduced the parameter $\Lambda = R/6$, with $R$ the
scalar curvature of the invariant metric associated to $\eta$ on the
homogeneous spacetime with Klein pair $(\g,\h)$ with $\h$ the span of
$J_{AB}$.  Let us define $J_A = -\tfrac12 \epsilon_{ABC} J^{BC}$, where
we raise and lower indices with $\eta$ and where the Levi-Civita
symbol $\epsilon_{ABC}$ has been normalised to $\epsilon_{012}=1$.
Notice that this implies that $\epsilon^{012} =-1$ and hence that
\begin{equation}
  \epsilon_{ABC} \epsilon^{CDE} = -\delta_A^D \delta_B^E + \delta_A^E
  \delta_B^D  \qquad\text{and}\qquad \tfrac12 \epsilon_{ABC}
  \epsilon^{BCD} = - \delta_A^D.
\end{equation}
It therefore follows that $J_{AB} = \epsilon_{ABC} J^C$.

In terms of $J_A,P_A$ the Lie brackets in
equation~\eqref{eq:Lie-algebra} are given by
\begin{equation}
  \label{eq:Lie-algebra-new-basis}
  \begin{split}
    [J_A, J_B] &= \epsilon_{ABC} J^C\\
    [J_A, P_B] &= \epsilon_{ABC} P^C\\
    [P_A, P_B] &= - \Lambda \epsilon_{ABC} J^C.
  \end{split}
\end{equation}

Let $\h$ denote the span of the $J_A$ and let the complement $\fm$ be
spanned by the $P_A$.  Then as shown in
\cite[§4.1.3]{Figueroa-OFarrill:2019sex}, the only $\h$-equivariant
bilinear maps $N \colon \fm \times \fm \to \fm$ are of the form
\begin{equation}
  \label{eq:Nomizu}
  N(P_A, P_B) = t \epsilon_{ABC} P^C,
\end{equation}
for some real number $t$.  The invariant connection corresponding to
such a Nomizu map has torsion
\begin{equation}
  \label{eq:torsion}
  T(P_A, P_B) = N(P_A,P_B)-N(P_B,P_A) - [P_A,P_B]_{\fm} = 2 t \epsilon_{ABC} P^C
\end{equation}
and curvature
\begin{equation}
  \label{eq:curvature}
  \begin{split}
      R(P_A, P_B)P_C &= N(P_A,N(P_B,P_C)) -  N(P_B,N(P_A,P_C)) \\
      &\quad - N([P_A,P_B]_{\fm},P_C) -  [[P_A,P_B]_\h, P_C] \\
      &= - (\Lambda + t^2) \left( \eta_{CA} P_B - \eta_{CB} P_A \right).
  \end{split}
\end{equation}

\subsection{The Lie algebras}
\label{sec:MB-algebras}

A natural question now arises: is there a Lie algebra structure on the
span of $J_A, P_A$ such that its canonical invariant connection (the
one corresponding to a zero Nomizu map) agrees with the invariant
connection just described?

Letting $[-,-]'$ denote the new Lie bracket, it follows from the
expression~\eqref{eq:Torsion-Curvature-Canonical} for the torsion and
curvature of the canonical connection that
\begin{equation}
  T(P_A,P_B) = - [P_A,P_B]'_\fm  \qquad\text{and}\qquad R(P_A,P_B)P_c = - [[P_A,P_B]'_\h, P_c]',
\end{equation}
so that taking $T$ and $R$ as in equations~\eqref{eq:torsion} and
\eqref{eq:curvature}, respectively, this can be achieved by defining
\begin{equation}
  [P_A, P_B]'_\fm = -2t \epsilon_{ABC} P^C
\end{equation}
and
\begin{equation}
  [P_A, P_B]'_\h = - (\Lambda + t^2) \epsilon_{ABC} J^C.
\end{equation}
In summary, defining the Lie algebra $\g'$ spanned by $J_A, P_A$ with brackets
\begin{equation}
  \label{eq:new-algebra}
  \begin{split}
    [J_A, J_B]' &= \epsilon_{ABC} J^C\\
    [J_A, P_B]' &= \epsilon_{ABC} P^C\\
    [P_A, P_B]' &= - (\Lambda + t^2) \epsilon_{ABC} J^C - 2t \epsilon_{ABC} P^C,
  \end{split}
\end{equation}
and letting $\h$ still denote the span of the $J_A$, the Klein pairs
$(\g',\h)$ define reductive homogeneous spaces whose canonical
invariant connections have torsion and curvature given by equations
\eqref{eq:torsion} and \eqref{eq:curvature}, respectively.

It is convenient to introduce parameters $p:= - (\Lambda + t^2)$ and
$q:= - 2t$, so that the brackets now become (dropping primes)
\begin{equation}
  \label{eq:MB-algebra}
  \begin{split}
    [J_A, J_B] &= \epsilon_{ABC} J^C\\
    [J_A, P_B] &= \epsilon_{ABC} P^C\\
    [P_A, P_B] &= \epsilon_{ABC} (p J^C + q P^C),
  \end{split}
\end{equation}
a Lie algebra first written in \cite{Geiller:2020edh} and which was
inspired by topological theories of gravity with torsion
\cite{Mielke:1991nn} and  their Chern--Simons formulation
\cite{Blagojevic:2002,BlagojevicCvetkovic:2004,Giacomini:2006dr}.

We emphasise that the parameters $q$ and $p$ are related to the
torsion and curvature of the canonical connection on the homogeneous
Klein pair $(\g,\h)$, with $\g$ the Lie algebra with
brackets~\eqref{eq:MB-algebra} and $\h$ the span of the $J_A$.  If the
torsion is different from zero, then the canonical connection is not
the Levi-Civita connection of any invariant metric and hence $p$ will
\emph{not} be the scalar curvature of any invariant metric.  Indeed,
it is perfectly possible for a homogeneous space $(\g,\h)$ to admit a
flat invariant connection with torsion and at the same a unique (up to
homothety) invariant metric whose Levi-Civita connection is not flat.
In the present context this will be the case with $\AdS_3$.

We should remark that the Lie algebra defined by~\eqref{eq:MB-algebra}
is, of course, isomorphic to that in
equation~\eqref{eq:Lie-algebra-new-basis}, simply via the invertible
linear transformation $J_A \mapsto J_A$ and
$P_A \mapsto P_A - t J_A = P_A - \tfrac12 q J_A$.  In other words, if
$P_A$ obey the bracket in equation~\eqref{eq:MB-algebra}, then
$P_A - \tfrac12 q J_A$ obey the bracket in
equation~\eqref{eq:Lie-algebra-new-basis} with
$\Lambda = - (p + \tfrac14 q^2)$.  This shows that different invariant
connections correspond to different choices of reductive splittings
$\g = \h \oplus \fm$; equivalently, different choices of the
complementary subspace\footnote{In the context of a chosen geometric
realisation $G/H$ of the Klein pair $(\g,\h)$ with the projection
$\pi : G\to G/H$ defining a principal $H$-bundle, then different
invariant connections correspond to different choices of an
$H$-invariant horizontal sub-bundle $\mathcal{H} \subset TG$
complementary to the vertical sub-bundle $\mathcal{V} = \ker \pi_*
\subset TG$.} $\fm$.

The parameter $\Lambda$ in the Lie algebra acts as a discriminant of
the different isomorphism types.  If $\Lambda < 0$, the Lie algebra is
isomorphic to $\so(2,2)$ and this decomposes into two commuting simple
subalgebras.  A more general statement is that the span of
\begin{equation}
  L_A = P_A + \lambda J_A
\end{equation}
is a Lie subalgebra provided that
\begin{equation}
  \lambda^2 + q\lambda - p = 0,
\end{equation}
with roots
\begin{equation}
  \lambda_\pm = \frac{-q \pm \sqrt{q^2+4p}}{2}.
\end{equation}
If $q^2+4p\geq 0$, which corresponds to $\Lambda \leq 0$, the roots
are real.  If $q^2 + 4p > 0$, then $\Lambda < 0$ and the homogeneous
spacetime is anti~de~Sitter.  The Lie algebra is isomorphic to
$\so(2,2)$ and because there are two different roots, we have a
decomposition
\begin{equation}
  \so(2,2) \cong \so(2,1) \oplus \so(2,1).
\end{equation}
If $q^2 + 4 p =0$, there is a double root and we have an abelian ideal
spanned by $P_A - \tfrac{q}{2} J_A$, corresponding to the Poincaré
algebra.  Finally, if $q^2+4p<0$, the roots are complex and there is
no such real decomposition.  This is the case of de~Sitter spacetime,
whose isometry Lie algebra is isomorphic to $\mathfrak{so}(3,1)$.

\subsection{Central extensions}
\label{sec:central-extensions}

In this section we discuss (the lack of) central extensions of the Lie
algebra $\g$ given by equation~\eqref{eq:MB-algebra}.  Central
extensions of $\g$ are classified up to
equivalence\footnote{Equivalence of central extensions is a
  refinement of the notion of Lie algebra isomorphism: it is an
  isomorphism which acts as the identity on the Lie algebra being
  extended.} by the second Chevalley--Eilenberg cohomology $H^2(\g)$.
The differential of the Chevalley--Eilenberg complex is obtained by
dualising the Lie bracket in \eqref{eq:MB-algebra} and extending as an
odd derivation.

Canonically dual to the basis $J_A, P_A$ for $\g$, we have the basis
$\lambda^A, \pi^A$ for $\g^*$.  The Lie bracket defines a linear map
$[-,-] \colon \ext{2} \g \to \g$ whose transpose $d \colon \g^* \to
\ext{2}\g^*$ is given in terms of this dual basis by
\begin{equation}
  \label{eq:CE-differential}
  \begin{split}
    d \lambda_A &= -\tfrac12 \epsilon_{ABC} \lambda^B \wedge \lambda^C - \tfrac12 p \epsilon_{ABC} \pi^B \wedge \pi^C\\
    d \pi_A &= - \epsilon_{ABC} \lambda^B \wedge \pi^C - \tfrac12 q \epsilon_{ABC} \pi^B \wedge \pi^C,
  \end{split}
\end{equation}
where we lower indices with $\eta_{AB}$.  We then extend $d$ as an odd
derivation over the wedge product to define the Chevalley--Eilenberg
complex $d \colon \ext{n}\g^* \to \ext{n+1}\g^*$.  The Lie algebra
$\g$ acts on the Chevalley--Eilenberg complex via the algebraic
version of the Lie derivative: $L_X = d \imath_X + \imath_X d$, where
$\imath_X \colon \ext{n}\g^* \to \ext{n-1} \g^*$ is contraction by
$X \in \g$.  It is clear from this expression that $d L_X = L_X d$ and
hence this action sends cocycles (kernel of $d$) to coboundaries
(image of $d$) and thus acts trivially on the cohomology.  The
subalgebra $\h \subset \g$ spanned by $J_A$ is simple and hence acts
completely reducibly on any finite-dimensional representation.  This
means that the Chevalley--Eilenberg cohomology $H^\bullet(\g)$ can be
calculated from the much smaller subcomplex of $\h$-invariant forms.
There are no $\h$-invariant elements in $\g^*$, but there is a
one-dimensional subspace of $\h$-invariant two-forms in $\ext{2}\g^*$,
spanned by
\begin{equation}
  \varphi = \eta_{AB} \lambda^A \wedge \pi^B.
\end{equation}
A calculation using equation~\eqref{eq:CE-differential} shows that
\begin{align*}
    d \varphi &= \eta_{AB} d\lambda^A \wedge \pi^B - \eta_{AB} \lambda^A \wedge d\pi^B\\
    &= d \lambda_A \wedge \pi^A - \lambda^A \wedge d\pi_A\\
    &= (-\tfrac12 \epsilon_{ABC} \lambda^B \wedge \lambda^C - \tfrac12
    p \epsilon_{ABC} \pi^B \wedge \pi^C) \wedge \pi^A - \lambda^A
    \wedge (- \epsilon_{ABC} \lambda^B \wedge \pi^C - \tfrac12 q \epsilon_{ABC} \pi^B \wedge \pi^C)\\
    &= \tfrac12 \epsilon_{ABC} \pi^A \wedge \lambda^B \wedge
      \lambda^C  - \tfrac12 p \epsilon_{ABC} \pi^A \wedge \pi^B \wedge
      \pi^C + \tfrac12 q \epsilon_{ABC} \lambda^A \wedge \pi^B \wedge \pi^C,
\end{align*}
which is nonzero for any value of $p$ and $q$.  We conclude that there
are no $\h$-invariant $2$-cocycles and hence $H^2(\g) = 0$.

We remark that we could have restricted our calculation to the case
where $4 p + q^2 = 0$ from the start, since otherwise the Lie algebra
$\g$ is semisimple and hence the second Whitehead Lemma already says
that $H^2(\g) =0$.  That case is isomorphic to the Poincaré algebra,
where $p=q=0$.  We would have then had to calculate as we did above to
deduce that $H^2(\g) = 0$ for $\g$ the Poincaré algebra.

\subsection{Quadratic Casimirs}
\label{sec:quadratic-casimirs}

In this section we determine the quadratic Casimirs of the Lie
algebra~\eqref{eq:MB-algebra}, recovering results of
\cite{Geiller:2020edh} (compare their equation (2.13) with
\eqref{eq:casimirs} below).

We consider a general Lorentz-invariant quadratic element in the
universal enveloping algebra
\begin{equation}
  \begin{split}
    C &= \alpha\,\eta^{AB}J_AJ_B + \beta\,\eta^{AB}J_AP_B +  \gamma\,\eta^{AB}P_AP_B\\
    &= \alpha J^2 + \beta J \cdot P + \gamma P^2.
  \end{split}
\end{equation}
Since $\eta$ is Lorentz-invariant, each term in $C$ is separately
invariant under the adjoint action of $J_A$.  It therefore remains to
impose ad-invariance under $P_C$: $\ad_{P_C} C = 0$.  Computing in the
universal enveloping algebra, one finds
\begin{equation}
  \begin{split}
    \mathrm{ad}_{P_C}(J^2) &= 2\left(\epsilon_C^{\ AB}J_AP_B + P_C\right),\\
    \mathrm{ad}_{P_C}(J\cdot P) &= q\,\left(\epsilon_C^{\ AB}J_AP_B + P_C\right),\\
    \mathrm{ad}_{P_C}(P^2) &= -2p\,\left(\epsilon_C^{\ AB}J_AP_B + P_C\right).
  \end{split}
\end{equation}
Therefore
\begin{equation}
  \mathrm{ad}_{P_C}(C) = (2\alpha + \beta q - 2\gamma p) \left(\epsilon_C^{\ AB}J_AP_B + P_C\right),
\end{equation}
so that $C$ is a Casimir if and only if
\begin{equation}
2\alpha + \beta q - 2\gamma p = 0,\qquad\Rightarrow\qquad\alpha = p\gamma - \tfrac{q}{2}\beta.
\end{equation}

In summary, there is a two-dimensional space of quadratic casimirs,
with natural basis
\begin{equation}\label{eq:casimirs}
  C_1 = \eta^{AB} J_A (P_B - \tfrac{q}2 J_B) \qquad\text{and}\qquad
  C_2 = \eta^{AB} P_A P_B + p \eta^{AB} J_A J_B.
\end{equation}
It is perhaps remarkable that for $p=0$, $\eta^{AB} P_A P_B$ is a
casimir, so that not just for Minkowski spacetime but also for
anti~de~Sitter there is an invariant notion of mass.

\subsection{Flat invariant connections}
\label{sec:flat-inv-conns}

Another natural question is whether there exists a flat invariant
connection.  From the expression \eqref{eq:curvature} for the
curvature, we see that this happens if and only if
$\Lambda = - t^2 \leq 0$.  The case $t=0$ is Minkowski spacetime with
the Levi-Civita connection, which is both torsion-free and flat.  If
$t\neq 0$, there are two solutions $t = \pm \sqrt{-\Lambda}$, which
requires $\Lambda < 0$, corresponding to anti~de~Sitter spacetime.

The existence of these two flat invariant connections can be explained
by the fact that three-dimensional anti~de~Sitter spacetime is locally
isometric to the Lie group $\SL(2,\RR)$ relative to a bi-invariant
lorentzian metric agreeing at the identity with a multiple of the
Killing form.  As shown by Cartan and Schouten in
\cite{CartanSchouten1} initially in the semisimple case, Lie groups
admit two natural invariant flat connections $\nabla^{(\pm)}$,
associated to the two trivialisations of the tangent bundle by left-
and right-invariant vector fields.  In other words,
$\nabla^{(\pm)} X = 0$ if and only if $X$ is left/right-invariant.

What about de Sitter spacetime? A simply-connected manifold $M$ admits
a flat affine connection $\nabla$ if and only if it is parallelisable;
that is, if and only if its tangent bundle is trivial. This means that
there is a global frame $e_a$ for $TM$, which is parallel under
$\nabla$.  Conversely, given a global frame $e_a$ for $TM$, we define
a flat affine connection $\nabla$ by the requirement that
$\nabla e_a = 0$.  As every three-dimensional orientable manifold has
trivial tangent bundle, three-dimensional de~Sitter spacetime is
parallelisable and hence it admits flat affine connections. What the
above analysis shows is that none of these flat affine connections are
invariant under $\SO(3,1)$.

\subsection{The homogeneous geometries}
\label{sec:homog-geom}

We now study the homogeneous geometries with Klein pairs $(\g,\h)$
where $\g$ is the Lie algebra in equation~\eqref{eq:MB-algebra} and
$\h \subset \g$ is the subalgebra spanned by the $J_A$.  We will
choose a Lie group $G$ with Lie algebra $\g$.  The Lie correspondence
assigns to the subalgebra $\h$ a connected subgroup $H \subset G$ and
we will let $M = G/H$ denote the corresponding homogeneous spacetime.
We will give coordinates $x^A$ to $M$ via the coset representative
$g(x) = \exp(x^A P_A)$.  Thinking of this as a map $g : M \to G$, this
allows us to pull back the left-invariant Maurer--Cartan one-form
$\vartheta \in \Omega^1(G;\g)$ to $M$, resulting in a $\g$-valued
one-form $g^*\vartheta \in \Omega^1(M;\g)$.

The left-invariant Maurer--Cartan one-form $\vartheta$ obeys the
structure equation
\begin{equation}
  \label{eq:MC-structure-equation}
  \dd\vartheta + \tfrac12 [\vartheta,\vartheta] = 0,
\end{equation}
where the second term is the Lie bracket in $\g$ and the wedge product
of one-forms.  In other words, if $X,Y$ are vector fields in the group
$G$,
\begin{equation}
  \label{eq:bracket}
  \tfrac12 [\vartheta,\vartheta](X,Y) = [\vartheta(X),\vartheta(Y)].
\end{equation}
Let us expand $\vartheta$ relative to the basis for $\g$:
\begin{equation}
  \label{eq:MC-one-form-expanded}
  \vartheta = \widetilde\omega^A J_A + \widetilde\theta^A P_A,
\end{equation}
where we have introduced the component one-forms $\widetilde\omega^A, \widetilde\theta^A
\in \Omega^1(G)$.  The structure
equation~\eqref{eq:MC-structure-equation} becomes
\begin{equation}
  \label{eq:MC-structure-equation-expanded-curvature}
  \dd \widetilde\omega_A + \tfrac12 \epsilon_{ABC} \widetilde\omega^B \wedge \widetilde\omega^C + \tfrac12 p\ \epsilon_{ABC} \widetilde\theta^B \wedge \widetilde\theta^C = 0
\end{equation}
and
\begin{equation}
  \label{eq:MC-structure-equation-expanded-torsion}
  \dd \widetilde\theta_A + \epsilon_{ABC} \widetilde\omega^B \wedge \widetilde\theta^C + \tfrac12 q\ \epsilon_{ABC} \widetilde\theta^B \wedge \widetilde\theta^C = 0.
\end{equation}

Pretending, for ease of exposition, that $G$ is a matrix
group\footnote{This is not the case for $G$ the simply-connected Lie
  group with Lie algebra $\so(2,2)$, for instance.} we write
\begin{equation}
  g^*\vartheta = g^{-1} \dd g = \exp(-x\cdot P) \dd \exp(x \cdot P).
\end{equation}
This can be calculated from the formula for the derivative of the
exponential map (see, e.g., \cite[§1.2, Theorem~5]{MR1889121}), resulting
in
\begin{equation}
  \label{eq:MC-pullback-formula}
  g^*\vartheta = g^{-1} \dd g = D(\ad_{x\cdot P})(\dd x \cdot P),
\end{equation}
where the holomorphic function $D(z) = \frac{1-e^{-z}}{z}$ is defined
via its power series expansion
\begin{equation}
  \label{eq:D-power-series}
  D(z) = \sum_{n=0}^\infty \frac{(-1)^n z^n}{(n+1)!}
\end{equation}
and the linear map $\ad_{x\cdot P} : \g \to \g$ is defined by
$\ad_{x\cdot P} = [x^A P_A , -]$.

There are two extreme cases where the calculations simplify: $q=0$,
which corresponds to the Levi-Civita connection of the invariant
metric, and $p=0$, which for Minkowski and anti~de~Sitter spacetimes,
corresponds to a flat connection.  We can treat both of these extreme
cases simultaneously by demanding only that $pq = 0$.  Indeed, if $pq
= 0$, then a calculation shows that
\begin{equation}
  \ad^3_{x\cdot P} (\dd x \cdot P) = x^2 (p + q^2) \ad_{x \cdot P} (\dd x
  \cdot P),
\end{equation}
where $x^2 = \eta_{AB}x^A x^B$.  Iterating we see that
\begin{equation}
  \begin{split}
    \ad^{2k+1}_{x \cdot P} (\dd x \cdot P) &= x^{2k} (p+q^2)^k \ad_{x\cdot P} (\dd x \cdot P)\\
    \ad^{2k+2}_{x \cdot P} (\dd x \cdot P) &= x^{2k} (p+q^2)^k \ad^2_{x\cdot P} (\dd x \cdot P),
  \end{split}
\end{equation}
where\footnote{For $p,q$ general, $\ad^2_{x\cdot P} (\dd x \cdot P)$
  gets a term proportional to $pq J$, and one does not have the above
  nice recursion for the even and odd powers of $\ad_{x \cdot
    P}$. Since we are mainly interested in the $p=0$ case, we present
  this particular result to keep the expressions as simple as
  possible.}
\begin{equation}
  \label{eq:ad-and-ad2}
  \begin{split}
      \ad_{x\cdot P} (\dd x \cdot P) &= x^A \dd x^B \epsilon_{ABC} \left( p J^C + q P^C \right)\\
      \ad^2_{x\cdot P} (\dd x \cdot P) &= x^2 (p+q^2) \dd x \cdot P - (p + q^2) x \cdot \dd x\, x \cdot P.
  \end{split}
\end{equation}

Inserting this into equation~\eqref{eq:MC-pullback-formula} and using
equation~\eqref{eq:D-power-series}, we see that
\begin{equation}
  \begin{split}
    D(\ad_{x\cdot P})(\dd x \cdot P) &= \dd x \cdot P - \sum_{k=0}^\infty  \frac{x^{2k}(p+q^2)^k}{(2k+2)!} \ad_{x \cdot P} (\dd x \cdot P) +    \sum_{k=0}^\infty \frac{x^{2k}(p+q^2)^k}{(2k+3)!} \ad^2_{x \cdot P} (\dd x \cdot P)\\
    &= \dd x \cdot P + C(x) \ad_{x\cdot P} (\dd x \cdot P) + \frac{S(x)-1}{(p+q^2)x^2} \ad^2_{x \cdot P} (\dd x \cdot P),
  \end{split}
\end{equation}
where we have introduced the shorthands
\begin{equation}
  C(x) = \frac{1-\cosh\sqrt{(p+q^2)x^2}}{(p+q^2)x^2}  \qquad\text{and}\qquad S(x) = \frac{\sinh\sqrt{(p+q^2)x^2}}{\sqrt{(p+q^2)x^2}}. 
  \label{eq:CS1}
\end{equation}
These functions $C(x)$ and $S(x)$ are defined unambiguously by their
power series expansions.  Nevertheless, we have  chosen to
abbreviate the power series in terms of hyperbolic functions with the
tacit assumption that $(p+q^2)x^2>0$.  It may of course be the case
that this number is either zero or negative.  If zero, then $C(x) =
-\tfrac12$ and $S(x) = 1$; whereas if negative, then it would be more
appropriate to abbreviate the power series with trigonometric
functions
\begin{equation}
  C(x) = \frac{1-\cos\sqrt{-(p+q^2)x^2}}{(p+q^2)x^2}
  \qquad\text{and}\qquad S(x)=\frac{\sin\sqrt{-(p +q^2)x^2}}{\sqrt{-(p+q^2)x^2}}.
   \label{eq:CS2}
\end{equation}

Using equations~\eqref{eq:ad-and-ad2}, we arrive at
\begin{equation}
  g^* \vartheta = S(x) \dd x \cdot P + \frac{1 - S(x)}{x^2} x \cdot \dd x \, x \cdot P + C(x) x^A \dd x^B \epsilon_{ABC} (p J^C + q P^C).
\end{equation}

We now decompose the $\g$-valued one-form $g^*\vartheta$ according to $\g = \h \oplus \fm$ into $g^*\vartheta = \theta^A P_A + \omega^A J_A$, where
\begin{equation}
  \label{eq:theta-and-omega}
  \begin{split}
    \theta^A &= S(x) \dd x^A + \frac{1-S(x)}{x^2} x^A x \cdot \dd x + q C(x) \epsilon^A{}_{BC} x^B \dd x^C\\
    \omega^A &= p C(x) \epsilon^A{}_{BC} x^B \dd x^C.
 \end{split}
\end{equation}
Therefore if $q=0$, then $\theta^A = S(x) \dd x^A + \frac{1-S(x)}{x^2} x^A x \cdot \dd x$ and the torsion vanishes:
\begin{equation}
  \label{eq:Theta}
  \Theta_A = \dd\theta_A + \epsilon_{ABC} \omega^B \wedge \theta^C = - \frac{q}{2} \epsilon_{ABC} \theta^B \wedge \theta^C = 0,
\end{equation}
whereas if $p=0$, we see that $\omega^A = 0$ and using the structure
equation~\eqref{eq:MC-structure-equation-expanded-curvature} we see that
the curvature vanishes:
\begin{equation}
  \label{eq:Omega}
  \Omega_A = \dd\omega_A + \tfrac12 \epsilon_{ABC} \omega^B \wedge \omega^C = -\tfrac12 p \epsilon_{ABC} \theta^B \wedge \theta^C = 0,
\end{equation}
and (\ref{eq:MC-structure-equation-expanded-torsion}) becomes
\begin{equation}
	\dd\theta^A + \frac{q}{2} \epsilon^A{}_{BC}\theta^B\wedge\theta^C =0.
	\label{eq:MC}
\end{equation}
In this case (i.e., $p=0$), using $\theta^A = \theta^A{}_M \dd x^M$ with
\begin{equation}
	\theta^A{}_M
	=
	S(x)\delta^A{}_M
	+
	\frac{1-S(x)}{x^2}x^A x_M
	+
	qC(x)\epsilon^A{}_{BM}x^B,
	\label{eq:app-frame-coframe-components}
\end{equation}
 the invariant metric $G = \eta_{AB}\theta^A \theta^B$ is given by
\begin{equation}
  G = \left( S^2 -q^2 C^2 x^2\right)\,\dd x^2 + \left(1 - S^2 +q^2 C^2
    x^2 \right) \frac{(x\cdot \dd x)^2}{x^2},
\end{equation}
which can be written as $G = G_{MN} \dd x^M \dd x^N$, where
\begin{equation}
  G_{MN} = F(x)\,\eta_{MN} + \left(1-F(x)\right)\frac{x_M x_N}{x^2},
  \label{eq:gMB}
\end{equation}
where we have introduced
\begin{equation}
  F(x):=S^2(x)-q^2C^2(x)x^2,
  \label{eq:F}
\end{equation}
which has different specific forms depending on the sign of $x^2$ (see (\ref{eq:CS1}) and (\ref{eq:CS2}) for $p=0$),
\begin{equation}
F(x)=\begin{cases}
	\left(\frac{2}{\sqrt{q^2x^2}}\sinh\frac{\sqrt{q^2x^2}}{2}\right)^2, & x^2>0,\\
	     \left(\frac{2}{\sqrt{-q^2x^2}}\sin\frac{\sqrt{-q^2x^2}}{2}\right)^2,                                                                                & x^2<0,                          
\end{cases}
\label{eq:Fcases}
\end{equation}
while $F(x)=1$ if $x^2=0$. Notice that for $x^2<0$ there are discrete points for which $F=0$. At those values the exponential-coordinate coframe ceases to be invertible, so the present chart does not cover the homogeneous geometry; the metric itself remains nondegenerate in a regular chart.
 
This metric is of course maximally symmetric and hence, in particular,
Einstein.  A direct computation of the Ricci curvature (of the
Levi-Civita connection) yields
\begin{equation}
  R_{MN} = -\frac{q^2}{2} G_{MN}, \qquad R = -\frac{3}{2} q^2.
\end{equation}
Thus if $q\neq 0$, the spacetime is $\mathrm{AdS}_3$ with cosmological
constant
\begin{equation}
  \Lambda = -\frac{q^2}{4} = \frac16 R,
\end{equation}
whereas if $q = 0$ then it is Minkowski spacetime. 

One can see that the inverse frame components, defined by
\begin{equation}
	e_A{}^M\theta^B{}_M=\delta_A{}^B,
	\qquad
	e_A{}^M\theta^A{}_N=\delta^M{}_N,
	\label{eq:app-frame-inverse-def}
\end{equation}
are given by
\begin{equation}
	e_A{}^M
	=
	D(x)\delta_A{}^M
	+
	\bigl(1-D(x)\bigr)
	\frac{x_Ax^M}{x^2}
	-
	\frac q2\, \epsilon_A{}^{BM}x_B,
	\label{eq:app-frame-inverse-components2}
\end{equation}
with
\begin{equation}
	D(x)= -\frac{1}{2}\frac{S(x)}{C(x)} = \begin{cases}
		\frac{\sqrt{q^2x^2}}{2}\coth\frac{\sqrt{q^2x^2}}{2}, & x^2>0,\\
		\frac{\sqrt{-q^2x^2}}{2}\cot\frac{\sqrt{-q^2x^2}}{2}, & x^2<0,
	\end{cases}
	\label{eq:app-frame-Ccal-def}
\end{equation}
with $D(x)=1$ if $x^2=0$, and assuming $F(x)\neq 0$.

Finally, the inverse of the space-time metric is
\begin{equation}
	G^{MN}
	=
	\eta^{AB}e_A{}^M e_B{}^N
	=
	\frac{1}{F(x)}\eta^{MN}
	+
	\left(1-\frac{1}{F(x)}\right)
	\frac{x^Mx^N}{x^2}.
	\label{eq:app-frame-inverse-metric}
\end{equation}

\subsubsection*{Conformally flat coordinates}

The MB metric is conformally flat, and this can be made explicit by introducing
new spacetime coordinates. In the spacelike patch $x^2>0$ we set $\rho=\sqrt{x^2}$ and define
\begin{equation}
	y^M = r(\rho) k^M ,
	\qquad
	x^M = \rho k^M ,
	\qquad
	\eta_{MN}k^M k^N = 1 ,
\end{equation}
where
\begin{equation}
	r(\rho)=\frac{2}{q}\tanh\left(\frac{q\rho}{4}\right).
\end{equation}
The null hypersurface is not covered by this chart, and the timelike patch requires the corresponding analytic continuation.

Starting from\footnote{The notation \(\dd k^2\) does not represent a positive-definite spherical metric.  It is the metric induced by \(\eta_{MN}\) on the hypersurface \(k^2=\varepsilon\).  In the spacelike region \(x^2>0\), one has \(\varepsilon=+1\), and therefore
	\begin{equation}
		\dd s^2
		=
		\dd\rho^2+R^2(\rho)\dd s_{\mathrm{dS}}^2,
	\end{equation}
	where the unit pseudo-sphere \(k^2=1\) is a de Sitter hypersurface and its induced metric \(\dd s_{\mathrm{dS}}^2\) is Lorentzian.  Thus the Lorentzian direction is contained in the angular part of the metric.  In the timelike region \(x^2<0\), one has \(\varepsilon=-1\), so that
	\begin{equation}
		\dd s^2
		=
		-\dd\rho^2+R^2(\rho)\dd s_{\mathbb H}^2,
	\end{equation}
	where \(k^2=-1\) is a hyperbolic hypersurface with a positive-definite induced metric and the radial coordinate itself is timelike.  Consequently, the full geometry is pseudo-riemannian in both regions.  The abbreviated expression
	\begin{equation}
		\dd s^2=\dd\rho^2+R^2(\rho)\dd k^2
	\end{equation}
	is correct only in the spacelike patch \(x^2>0\), with the essential understanding that \(\dd k^2\) is Lorentzian.  The null hypersurface \(x^2=0\) is not covered by this radial coordinate system and must be treated in a separate chart.
}
\begin{equation}
	\dd s^2
	=
	G_{MN}(x)\dd x^M\dd x^N
	=
	\dd\rho^2+R^2(\rho)\dd k^2,
\end{equation}
with
\begin{equation}
	R(\rho)=\frac{2}{q}\sinh\left(\frac{q\rho}{2}\right),
\end{equation}
we look for a conformal factor \(\Omega(\rho)\) such that
\begin{equation}
  \dd s^2
  =
  \Omega^2(\rho)\eta_{MN}\dd y^M\dd y^N
  =
  \Omega^2(\rho)\left(\dd r^2+r^2\dd k^2\right).
\end{equation}
Matching the angular terms gives
\begin{equation}
	\Omega(\rho)=\frac{R(\rho)}{r(\rho)}
	=
	2\cosh^2\left(\frac{q\rho}{4}\right),
\end{equation}
and using $y^2:=\eta_{MN}y^M y^N=r^2$
one can also write
\begin{equation}
	\Omega(y)=
	\frac{2}{1-\frac{q^2}{4}y^2}.
\end{equation}
Therefore the MB metric takes the manifestly conformally flat form
\begin{equation}
	\dd s^2
	=
	\frac{4}{\left(1-\frac{q^2}{4}y^2\right)^2}
	\eta_{MN}\dd y^M\dd y^N .
\end{equation}

\section{Massive particle dynamics in Mielke--Baekler spacetimes}
\label{sec:dyn-MB}

We now discuss particle dynamics in a Mielke--Baekler spacetime $M =
G/H$, paying particular attention to the teleparallel case $p=0$.  In
this section we discuss the massive particles and in
Section~\ref{sec:massless-particles} we discuss the massless particles.

\subsection{Particle actions from coadjoint orbits via nonlinear realisations}
\label{sec:particles-from-orbits}

The momentum of a particle propagating in a Mielke--Baekler spacetime
$M = G/H$ takes values in a homogeneous symplectic manifold of the
group $G$, which is known as the space of motions.  Since, as shown at
the Lie algebra level in Section~\ref{sec:central-extensions}, $G$
admits no nontrivial central extensions, such homogeneous symplectic
manifolds are (up to coverings) coadjoint orbits of $G$.

Let $\alpha \in \g^*$ and $\eO_\alpha$ the corresponding coadjoint
orbit.  Let $\pi_\alpha : G \to \eO_\alpha$ denote the orbit map
sending $g \in G$ to $\Ad^*_g \alpha$.  Let $G_\alpha\subset G$ be the
stabiliser of $\alpha$:
\begin{equation}
	G_\alpha = \left\{ g \in G ~\middle |~ \Ad^*_g \alpha = \alpha \right\}.
\end{equation}
Let $o \in M$ be a point with stabiliser $H$ and let $\varpi_o : G \to
M$ denote the orbit map sending $g \in G$ to $g \cdot o$.  The
homogeneous space of $G$ of smallest dimension which fibers over both
$M$ and $\eO_\alpha$ is the evolution space $\mathcal{E}$ of the
particle and it is given by the coset space $G/(G_\alpha \cap H)$.

Let $I \subset \RR$ be an interval and let $\gamma : I \to
\mathcal{E}$, sending $\tau \mapsto \gamma(\tau)$, be a curve in the
evolution space.  We can view this as a curve in the group by
composing with a coset representative $g : \mathcal{E} \to G$.  The
lagrangian is then given by pulling-back the left-invariant
Maurer--Cartan one-form on $G$ to the interval and contracting with
the moment $\alpha \in \g^*$, resulting in the action functional
\begin{equation}
	\label{eq:action_functional}
	S[\gamma] = \int_I \left<\alpha, (g \circ \gamma)^* \vartheta\right>.
\end{equation}
It is known that the extremals are curves in the evolution space
$\mathcal{E}$ whose velocities lie in the kernel of the presymplectic
structure on $\mathcal{E}$ defined by pulling back the
Kirillov--Kostant--Souriau symplectic form on $\eO_\alpha$ via the
projection $\mathcal{E} \to \eO_\alpha$.  These curves can then be
pushed down to the spacetime by the projection $\mathcal{E} \to M$
resulting in particle trajectories.

\subsection{Massive particle actions}
\label{sec:massive-particles}

We now describe massive particles in the Mielke--Baekler spacetimes.
We will take their momentum $\alpha \in \g^*$ to be such that
\begin{equation}
  \alpha(P_A) = m \eta_{A 0} \qquad\text{and}\qquad \alpha(J_A) = - s \eta_{A 0}.
\end{equation}
Equivalently, relative to the canonical dual basis $\lambda^A, \pi^A$
for $\g^*$, we can write $\alpha = - m \pi^0 + s \lambda^0$, where the
sign in the first term is such that the action functional in
equation~\eqref{eq:action_functional} corresponds, as we will see
below, to the standard particle lagrangian for a massive spinless
particle in Minkowski spacetime.

The quadratic Casimirs found in Section~\ref{sec:quadratic-casimirs}
define functions on $\g^*$ which are constant on coadjoint orbits: on
$\eO_\alpha$, $C_1 = \tfrac12 s(q s + 2 m)$ and
$C_2 = -(m^2 + p s^2)$, as can be seen by evaluating them at $\alpha$.

We remark on the curious fact, already mentioned in
\cite{Batlle:2014sca}, that although generically $\eO_\alpha$ is
$4$-dimensional, there exists a critical locus in the $(s,m)$ plane
for which $\eO_\alpha$ is $2$-dimensional.  In fact, it is easy to
determine the Lie-algebraic coadjoint action and from there the
dimension of the stabiliser $\g_\alpha \subset \g$, so that
$\dim \eO_\alpha = \dim \g - \dim \g_\alpha = 6 - \dim \g_\alpha$.
For any $X \in \g$ and $\alpha \in \g^*$, we have that
$\ad^*_X \alpha = - \alpha \circ \ad_X$, so that relative to the
canonical dual bases $J_A, P_A$ and $\lambda^A, \pi^A$ for $\g$ and
$\g^*$, respectively, we find
\begin{equation}\label{eq:coadjoint-action}
  \begin{aligned}
    \ad_{J_A}^* \lambda_B &=  \epsilon_{ABC} \lambda^C\\
    \ad_{P_A}^* \lambda_B &=  p \epsilon_{ABC} \pi^C\\
  \end{aligned}
  \qquad\qquad
  \begin{aligned}
    \ad_{J_A}^* \pi_B &= \epsilon_{ABC} \pi^C\\
    \ad_{P_A}^* \pi_B &=  \epsilon_{ABC} \left( \lambda^C + q \pi^C \right).\\
  \end{aligned}
\end{equation}

Taking $\alpha = - m \pi^0 + s \lambda^0$ and $X = a^A P_A + b^A J_A$,
\begin{equation}
	\ad^*_X \alpha = (- m a^1 + s b^1) \lambda^2 - (- m a^2 + s b^2)
	\lambda^1 + (- m q a^1 - m b^1 + s p a^1) \pi^2 - (- m q a^2 - m b^2 + s p a^2) \pi^1.
\end{equation}
Therefore $X \in \g_\alpha$ if and only if the following homogeneous
linear equation is satisfied
\begin{equation}
	\begin{pmatrix}
		-m & s & 0 & 0\\
		s p - m q& -m & 0 & 0 \\
		0 & 0 & -m & s\\
		0 & 0 & s p -m q & -m
	\end{pmatrix}
	\begin{pmatrix}
		a^1\\ b^1\\ a^2\\ b^2
	\end{pmatrix} =
	\begin{pmatrix}
		0 \\ 0 \\ 0 \\ 0
	\end{pmatrix}.
\end{equation}
The above matrix has determinant $(m^2 - p s^2 + m q s)^2$, which is
generically different from zero and hence $a^1 = a^2 = b^1 = b^2 = 0$.
This says that $\g_\alpha$ is spanned by $J_0, P_0$ and hence
$\dim \eO_\alpha = 6-2 = 4$.  However, if $m^2 = s (p s - m q)$, then
the rank of the matrix is equal to $2$ (if $s\neq 0$) or $0$ (if
$s = 0$, so that $m =0$ as well).  Since we are interested in
$\alpha \neq 0$, then we can take $s \neq 0$ in the critical case and
$\g_\alpha$ is now spanned by
$J_0, P_0, s P_1 + m J_1, s P_2 + m J_2$, so that $\dim \eO_\alpha = 6
-4 = 2$. The regular and critical cases implied by the above rank
condition have different dynamical consequences, which are analysed
for $p=0$, in Sections \ref{sec:massive-teleparallel} and
\ref{sec:massive-canonical}.

\subsection{Massive particles in the teleparallel Mielke--Baekler spacetime}
\label{sec:massive-teleparallel}

If $p=0$, which we may call the ``teleparallel'' case, $C_2 = -m^2$,
so that we may interpret $m$ as a mass.  Our choice of $\alpha$ then
describes the momentum of a massive spinning particle.

Let us write this more explicitly.  Since the evolution space breaks
the explicit Lorentz symmetry, it will be convenient to introduce a
different basis for the Lie algebra $\g$.  To this end we will define
generators, for $a \in \{1,2\}$, by
\begin{equation}
	\label{eq:kin-gens}
	J := -J_{12}, \quad H := P_0 \quad\text{and}\quad B_a = J_{0a},
\end{equation}
which together with $P_a$ span the Lie algebra.  In this basis, the
bracket \eqref{eq:MB-algebra} is given by
\begin{equation}
	\label{eq:MB-kin}
	\begin{aligned}
		[J,B_a] &= \epsilon_{ab} B_b\\
		[J,P_a] &= \epsilon_{ab} P_b\\
		[B_a, B_b] &= - \epsilon_{ab} J\\
		[B_a, P_b] &= \delta_{ab} H
	\end{aligned}
	\qquad\qquad
	\begin{aligned}
		[H,B_a] &= - P_a\\
		[H,P_a] &= p B_a + q \epsilon_{ab} P_b\\
		[P_a, P_b] &= -\epsilon_{ab} (p J + q H),\\
	\end{aligned}
\end{equation}
where $\epsilon_{ab}$ has been normalised to $\epsilon_{12} = 1$.
Indices are raised and lowered with $\delta_{ab}$, which often remains
tacit.  In terms of the basis of the Lie algebra in
equation~\eqref{eq:Lie-algebra-new-basis}, the relation is that $J =
J_0$ and $B_a = \epsilon_{ab} J_b$.  From now on we will only consider
the teleparallel case ($p=0$).

In the nonlinear realisation approach
\cite{Coleman:1969sm,Callan:1969sn,Volkov:1973vd} the coset
representative $g: \mathcal{E} \to G$ is given by
\begin{equation}\label{eq:repParticle}
  g(x,v)=g_0(x)\,b(v),
  \qquad
  g_0(x)=\exp(x^A P_A),
  \qquad
  b(v)=\exp(v^a B_a),
\end{equation}
where $v^a$ are the  Goldstone bosons associated to the broken boosts.
The pull-back to $\mathcal{E}$ of the left-invariant Maurer--Cartan
one-form is given by
\begin{equation}
	g^* \vartheta = g^{-1}\dd g=b^{-1} (g_0^{-1} \dd g_0) b + b^{-1}\dd b,
	\qquad\text{where}\qquad
	g_0^{-1}\dd g_0 = \theta^A P_A + \omega^A J_A,
\end{equation}
with $\theta^A$ and $\omega^A$ given by equation
\eqref{eq:theta-and-omega}.  Since we are in the teleparallel case,
$\omega^A = 0$.  Since the boost $b$ is a Lorentz transformation, its
adjoint representation is given by $\Lambda(v)\in SO(1,2)$ as follows:
\begin{equation}
	b^{-1}P_A b= P_B \Lambda(v)^B{}_A,
\end{equation}
where
\begin{equation}
	\Lambda(v)^A{}_B =\begin{pmatrix} \cosh v & v_b \frac{\sinh v}{v} \\ v^a \frac{\sinh v}{v} & \delta^a{}_b + \frac{\cosh v- 1}{v^2} v^a v_b 
	\end{pmatrix}\qquad\text{where}\qquad v=\sqrt{v^a v_ a}.
\end{equation}

Thus we write $g^* \vartheta = \vartheta^A P_A + \varpi^A J_A$, where
$\vartheta^A =\Lambda^A{}_B(v)\,\theta^B$ and $\varpi^A J_A = \varpi^a
B_a + \varpi_J J$ are defined by
\begin{equation}
	b^{-1}\dd b= \varpi^a B_a + \varpi_J J,
\end{equation}
with
\begin{equation}
	\varpi^a = \frac{\sinh v}{v}\,\dd v^a + \left(1-\frac{\sinh v}{v}\right) \frac{v^a v_b\,\dd v^b}{v^2},
\end{equation}
and
\begin{equation}
	\varpi_J = \frac{\cosh v-1}{v^2}\, \epsilon_{ab}v^a\,\dd v^b.
\end{equation}

The action functional~\eqref{eq:action_functional} for curves $\gamma
: I \to \mathcal{E}$ is then given explicitly by the integral of the
lagrangian one-form on the interval
\begin{equation}
	L\,\dd \tau = \left<- m \pi^0 + s \lambda^0, \gamma^*\vartheta^A P_A +
	\gamma^*\varpi^A J_A\right> = - m \gamma^* \vartheta^0 + s \gamma^* \varpi_J,
\end{equation}
where
\begin{equation}
	\gamma^*\vartheta^0 = \cosh v\, \gamma^*\theta^0 + \frac{\sinh v}{v}
	v_a \gamma^* \theta^a
\end{equation}
and
\begin{equation}
	\gamma^*\varpi_J = \frac{\cosh v - 1}{v^2} \epsilon_{ab} v^a \dot
	v^b \dd\tau,
\end{equation}
where we describe the curve $\gamma$ in terms of the coordinates as
$x^A(\tau), v^a(\tau)$ and where the dot denotes derivative with
respect to $\tau$.  It will be convenient to write the pull-back
$\gamma^*\theta^A$ of the coframe to the worldline of the particle as
$\gamma^*\theta^A = \theta^A_\tau  d\tau$, where $\theta^A_\tau =
\theta^A{}_M \dot x^M$.
In summary, the lagrangian is given by
\begin{equation}
  L = - m \left( \cosh v\, \theta_\tau^0 + \frac{\sinh v}{v} v_a \theta_\tau^a  \right) + s \left( \frac{\cosh v - 1}{v^2} \epsilon_{ab} v^a \dot v^b \right).
\end{equation}
It is convenient to introduce $\hat{v}^a = \frac{v^a}{v}$, in terms of
which
\begin{equation}\label{eq:Lteleparallel}
  L = - m \left( \cosh v\, \theta_\tau^0 + \sinh v \hat{v}_a \theta_\tau^a  \right) + s \left(\cosh v - 1\right) \epsilon_{ab} \hat{v}^a \dot{\hat{v}}^b.
\end{equation}

\subsubsection{Massive spinless particle and inverse Higgs constraint}
\label{sec:massive-spinless}

For $s=0$, the derivatives of the Goldstone variables $(v,\hat v^a)$ do not appear in the lagrangian and are auxiliary.  Varying \eqref{eq:Lteleparallel}
yields the constraints
\begin{align}
	\Pi^{ab}_\perp\, \theta^b_\tau &=0,\\
	\sinh v\, \theta^0_\tau + \cosh v\,\hat v^a \theta^a_\tau &= 0,
\end{align}
where $\Pi^{ab}_\perp :=  \delta^{ab} - \hat v^a\hat v^b$, which are
equivalent to the inverse Higgs constraint \cite{Ivanov:1975zq,McArthur:2010}
\begin{equation}\label{eq:IHC}
	\vartheta^a_\tau = 0.
\end{equation}
Solving it gives
\begin{equation}
	\theta^a_\tau = - \hat v^a\tanh v\,\theta^0_\tau.
\end{equation}
Substituting back reduces the action to the standard massive form
\begin{equation}\label{eq:NG}
  \left.S\right|_{p=0,s=0} = - m \int \dd\tau\,\sqrt{-\eta_{AB}\theta^A_\tau \theta^B_\tau} = -  m\int \dd\tau\,\sqrt{-G_{MN}(x)\dot x^M\dot x^N}.
\end{equation}

The resulting Euler--Lagrange equation is the geodesic equation for
the metric
\begin{equation}\label{eq:gMN}
  G_{MN}(x)=F(\rho)\eta_{MN}+\big(1-F(\rho)\big)\frac{x_Mx_N}{x^2},
\end{equation}
with $F$ given by (\ref{eq:F}).
Explicitly, these are
\begin{equation}
	\frac{d}{d\tau}\left(\frac{G_{MN}\dot x^{N}}{\sqrt{-G_{KL}\dot x^{K}\dot x^{L}}}\right)
	-\frac{1}{2}\frac{\partial_{M} G_{KL}\dot x^{K}\dot x^{L}}{\sqrt{-G_{RS}\dot x^{R}\dot x^{S}}}=0.
	\label{eq:p0_geodesic}
\end{equation}
In terms of momenta $p_M = m\frac{G_{MN}\dot x^{N}}{\sqrt{-G_{KL}\dot x^{K}\dot x^{L}}}$, the geodesic equation becomes
\begin{equation}\label{eq:LC_parallel_transport_p}
	\dot p_M - \widetilde\Gamma^N_{MP}\,\dot x^{P}p_N=0,\qquad\text{or equivalently}\qquad
	\widetilde{\nabla}_\tau p_M=0,
\end{equation}
where $\widetilde{\nabla}_\tau$ is the covariant derivative of the
pull-back of the Levi-Civita connection to the interval, whose
connection coefficients $\widetilde\Gamma^N{}_{MP}$ are given by the usual
Christoffel symbols
\begin{equation}\label{eq:LC_connection}
  \widetilde\Gamma^N_{MP} = \frac12\,G^{NQ}\left(\partial_M G_{QP}+\partial_P G_{QM}-\partial_Q G_{MP}\right).
\end{equation}

\subsubsection{Massive spinning particles}
\label{sec:massive-spinning}

We now turn on the spin.  For $s\neq 0$, the Goldstone boost becomes
dynamical. We denote the pull-back of the coframe to the worldline by
\begin{equation}
  E^A:=\theta^A_\tau=\theta^A{}_{M}(x)\dot x^M .
\end{equation}
Parametrising the Goldstone bosons as
\begin{equation}
v^a = v n^a, \quad n^an_a=1,\quad n^a = (\cos\varphi,\sin\varphi),
\end{equation}
one has $\epsilon_{ab} \hat{v}^a\dot{\hat{v}}^b = \dot\varphi$, and the lagrangian can be written as
\begin{equation}
	L=-m (\cosh v\, E^0 + \sinh v\, n_a E^a)+ s (\cosh v -1)\dot\varphi.
	\label{eq:c32-1}
\end{equation}
Using $\partial_\varphi n_a = -\epsilon_{ab} n^b$, the equations of motion for the Goldstone boson degrees of freedom $v,\varphi$ yield, respectively,
\begin{equation}\label{eq:c32}
  \begin{split}
    \dot\varphi &= \frac{m}{s} \frac{1}{\sinh v}\left( \sinh v\, E^0 + \cosh v\, n_a E^a\right)\\
    \dot v &= -\frac{m}{s}\epsilon_{ab}n^a E^b
  \end{split}
\end{equation}
where $v\neq 0$ has been assumed.

It is also useful to introduce the unit timelike Lorentz vector
\begin{equation}\label{eq:uDef}
	u^0:=-\cosh v,\qquad
	u^a:=\sinh v\,n^a,\qquad
	u_Au^A=-1,
\end{equation}
whose derivatives can be computed using equations~\eqref{eq:c32}:
\begin{align}
	\dot{u}^0 &= \frac{m}{s} \epsilon_{ab}u^a E^b = -\frac{m}{s} \epsilon^0{}_{ab} u^a E^b,\label{eq:c32-4}\\
	\dot{u}^a &= -\frac{m}{s} \epsilon^a{}_b u^b E^0 + \frac{m}{s} \epsilon^a{}_c u^0 E^c 
	          = -\frac{m}{s} \epsilon^a{}_{b0} u^b E^0 - \frac{m}{s} \epsilon^a{}_{0c} u^0 E^c
	          = -\frac{m}{s} \epsilon^a{}_{BC}u^B E^C,\label{eq:c32-5} 
\end{align} 
and can be put together in the conveniently covariant form
\begin{equation}
	\dot{u}^A = -\frac{m}{s} \epsilon^A{}_{BC} u^B E^C.
	\label{eq:c32-6}
\end{equation}
These are the equations of motion for the Goldstone bosons, which are coupled to the space-time degrees of freedom through $E^C= \theta^C{}_M (x) \dot x^M$.

In order to obtain the equations of motion for $x^M$ we rewrite the $m$ term in the lagrangian using the $u^A$ and the explicit form of $E^A$,
\begin{equation}
L= - m u_A \theta^A{}_M(x) \dot x^M + s (\cosh v -1)\dot\varphi,
\label{eq:c32-7}
\end{equation}	
from which one immediately obtains,
\begin{equation}
\dot{u}_A \theta^A{}_{M} = u_A (\partial_M \theta^A{}_{N}-\partial_N\theta^A{}_M)\dot{x}^N.
\label{eq:c32-8}
\end{equation}
The gradients of the dreibein components $\theta^A{}_M$ can be obtained from the Maurer-Cartan (\ref{eq:MC})
with $\theta^A = \theta^A{}_M \dd x^M$, and one gets
\begin{equation}
	\partial_N \theta^A{}_M - \partial_M \theta^A{}_N = q \epsilon^A{}_{BC} \theta^B{}_M \theta^C{}_N.
\label{eq:c32-9}
\end{equation}
Using this,  (\ref{eq:c32-8}) becomes
\begin{equation}
	\dot{u}_A \theta^A{}_{M} = q u_A  \epsilon^A{}_{BC}\theta^B{}_N\theta^C{}_M\dot{x}^N,
	\label{eq:c32-10}
\end{equation}
and contracting with the dual components $e^M{}_D$ one gets, after rearranging indices,
\begin{equation} 
	\dot{u}^A = q \epsilon^A{}_{BC} u^B E^C.
\label{eq:c32-11}
\end{equation}
Notice that this has the same form as the equation obtained from the
Euler-Lagrange equations (\ref{eq:c32-6}) for the Goldstone degrees of
freedom, and they are actually the same in the critical case
$m+qs=0$. This means that the theory exhibits different dynamical
sectors depending on whether or not $m+qs$ is zero.  This was also noticed for AdS$_3$ in
\cite{Batlle:2014sca} (see also \cite{Batlle:2025byv}), where it was
related to the existence of extremal black holes.

The equations of motion for $x^M$ can be written in terms of the spacetime momenta
\begin{equation}
  p_M = \frac{\partial L}{\partial \dot{x}^M} = - m u_A\theta^A{}_M
\end{equation}
as
\begin{equation}
  \dot{p}_M + m u_A \partial_M\theta^A{}_N \dot{x}^N = 0.
	\label{eq:c32-12}
\end{equation}
As shown in \eqref{eq:Weitz-conn-coeff}, the connection coefficients
of a Weitzenböck connection are defined in terms of the dual coframe $\theta^A{}_M$ as
\begin{equation}
  \Gamma_{MN}^P = e_A{}^P \partial_M\theta^A{}_N.
\end{equation}
One has then
\begin{equation}
  \Gamma_{MN}^P p_P = - m \partial_M \theta^A{}_N u_A
\end{equation}
and (\ref{eq:c32-12}) becomes
\begin{equation}\label{eq:c32-13}
  \dot{p}_M - \Gamma_{MN}^P \dot{x}^N p_P =0.
\end{equation}
This is not the parallel transport equation for a covector with
respect to the Weitzenböck connection with coefficients
$\Gamma_{MN}^P$, since the derivative index of the connection is not
the one contracted with $\dot x^N$. It would seem that it is the
parallel transport equation relative to a connection
whose connection coefficients $\overline\Gamma_{MN}^P:=
\Gamma_{NM}^P$.  These coefficients do define an affine connection
$\overline\nabla$, called the  transposed connection and defined
more invariantly by
\begin{equation}
  \overline\nabla_X Y := \nabla_Y X + [X,Y]
\end{equation}
or, equivalently by,
\begin{equation}
  \overline\nabla_X Y := \nabla_X Y - T(X,Y).
\end{equation}
It follows that $\overline\nabla$ has the opposite torsion to
$\nabla$: $\overline T(X,Y) = T(Y,X) = - T(X,Y)$.
The equation of motion for the spacetime momenta is then
\begin{equation}\label{eq:parallel-transport-transposed-connection}
  \overline\nabla_\tau p_M := \dot p_M - \overline\Gamma_{NM}^P\dot x^N p_P = 0.
\end{equation}

The transposed connection is a standard object in spaces with torsion,
where it is also called the associated connection; it appears, for
example, in the covariant formulation of Lie derivatives and
conservation laws in metric-affine geometry
\cite{ObukhovPortalesPuetzfeldRubilar2015, ObukhovRubilar2006,
  ObukhovRubilar2007}. Its relevance in particle dynamics is also
natural from the metric-affine perspective, where the canonical
energy-momentum current describes the transport of momentum and where
test-particle equations in non-riemannian backgrounds involve the full
connection, torsion and distortion tensors
\cite{PuetzfeldObukhov2014, IosifidisHehl2024}.

The transposed connection of a Weitzenböck connection need not be
again a Weitzenböck connection, but in the present context this turns
out to be the case.  The teleparallel geometry is either Minkowski
spacetime (if the torsion vanishes) or $\AdS_3$.  In the case of
Minkowski spacetime, the connection is symmetric and hence the
transposed connection agrees with the original connection (here, the
Levi-Civita connection).  In the case of $\AdS_3$, we have the happy
coincidence that $\AdS_3$ is locally isometric to the Lie group
$\SL(2,\RR)$ relative to a bi-invariant metric.  Every such Lie
group admits two Weitzenböck connections: the one relative to which
the left-invariant vector fields are parallel and the one relative to
which the right-invariant vector fields are parallel.  They are such
that their torsions are opposite.  Since the metric is bi-invariant,
both Weitzenböck connections are metric-compatible and such a
connection is uniquely determined by its torsion.  Therefore the
transpose of one of the Weitzenböck connections on $\AdS_3$, having
opposite torsion, must be the other Weitzenböck connection.

Let us now rewrite
equation~\eqref{eq:parallel-transport-transposed-connection} relative
to the Levi-Civita connection. Writing
\begin{equation}
	\Gamma_{MN}^P
	=
	\widetilde\Gamma_{MN}^P
	+
	K^P_{MN},
\end{equation}
where $\widetilde\Gamma_{MN}^P$ is the Levi-Civita connection and $K^P_{MN}$ is the contorsion, the transposed connection satisfies
\begin{equation}
	\overline\Gamma_{MN}^P
	=
	\widetilde\Gamma_{MN}^P
	+
	\overline K^P_{MN},
	\qquad
	\overline K^P_{MN}
	:=
	K^P_{NM}.
\end{equation}
Therefore
\begin{equation}
	\overline\nabla_\tau p_M=0
	\qquad\Longleftrightarrow\qquad
	\widetilde\nabla_\tau p_M
	=
	K^P_{MN}\dot x^N p_P .
\end{equation}
In the teleparallel MB background the torsion is totally antisymmetric in frame indices,
\begin{equation}
	T^A_{BC} = - q\,\epsilon^A{}_{BC},
\end{equation}
and the contorsion is
\begin{equation}
  \label{eq:contorsion}
	K^P_{MN}
	=
	- \tfrac q2\,e_A{}^P\epsilon^A{}_{BC}\theta^B{}_M\theta^C{}_N .
\end{equation}
Hence $K^P_{NM}=-K^P_{MN}$, so the transposed connection may equivalently be written as
\begin{equation}
	\overline\Gamma_{MN}^P
	=
	\widetilde\Gamma_{MN}^P
	-
	K^P_{MN}.
\end{equation}
This is consistent with the standard teleparallel interpretation, in
which the Levi-Civita form of the equations contains a contorsion, or
torsional force, term \cite{Aldrovandi:2013wha}. In the present
particle model this force term is sector-dependent: it vanishes in the
regular sector after imposing the constraints, whereas it survives in
the critical sector and can be rewritten in Papapetrou form.

These are the equations of motion in teleparallel form, without any
explicit forcing associated to the internal degrees of freedom given
by $u^A$ (although the space-time momenta $p$ still depend on the
Goldstone bosons $u$).

Using the explicit form of the contorsion in
equation~\eqref{eq:contorsion}, one finally writes
\begin{equation}
	\widetilde{\nabla}_\tau p_M = \frac{mq}{2} \epsilon_{ABC} u^A \theta^B{}_M E^C,
	\label{eq:c32-16}
\end{equation}
so that, using the ordinary Levi-Civita connection, the geodesic
equation seems to be modified by a  forcing
term. However, whether or not this term is actually zero depends on
the dynamical sector of the theory.

Indeed, combining (\ref{eq:c32-6}) and (\ref{eq:c32-11}) one has
\begin{equation}
	(m+qs)\epsilon^A{}_{BC} u^B E^C =0,
\end{equation}
and, if $m+qs\neq 0$,
\begin{equation}
  \epsilon^A{}_{BC} u^B E^C =0,
  \label{eq:c32-17}
\end{equation}
so that the RHS of equation~\eqref{eq:c32-16} vanishes, reducing to
\begin{equation}
  \widetilde{\nabla}_\tau p_M=0,
  \label{eq:c32-19}
\end{equation}
and no forcing term for the Levi-Civita connection appears in the
regular dynamical sector $m+qs\neq 0$.  Notice also that from
(\ref{eq:c32-11}) and (\ref{eq:c32-17})  it follows that
\begin{equation}
  \dot u^A =0,
  \label{eq:c32-19b}
\end{equation}
and thus the boost degrees of freedom are constant in the regular sector.
 
In the critical sector, defined by $m+qs=0$ , equation (\ref{eq:c32-17}) no longer holds, and solving for $E$ 
the single dynamical independent equation
\begin{equation}
\dot{u}^A = q \epsilon^A{}_{BC} u^B E^C
\end{equation}
one obtains
\begin{equation}
	E^A = -\frac{1}{q} \epsilon^A{}_{BC} u^B \dot{u}^C - (u\cdot E) u^A
	\label{eq:c32-20}
\end{equation}
showing the presence of a non parallel contribution. When inserted in
(\ref{eq:c32-16}) the parallel component drops, and using
$u\cdot\dot{u}=0$ the remaining term yields the equation
\begin{equation}
  \widetilde{\nabla}_\tau p_M= -\frac{m}{2}\theta^A{}_M \dot{u}_A =  \frac{qs}{2}\theta^A{}_M \dot{u}_A.
  \label{eq:c32-21}
\end{equation}

Hence, a forcing  term for the transport of the momenta using the ordinary Levi-Civita connection appears in the critical dynamical sector, but not in the regular one. This forcing term can be given a Papapetrou-like \cite{Papapetrou:1951} form by introducing  the spin tensor
\begin{equation}
	S^{AB} = -2s\epsilon^{ABC}u_C
	\label{eq:c32-22} 
\end{equation}
and rewriting everything in terms of $S^{AB}$ instead of $u^A$. We will first write down an equation for $\widetilde{\nabla}_\tau S^{AB}$ and then rewrite the right-hand side of (\ref{eq:c32-21}). In order to compute $\widetilde{\nabla}_\tau u^A$ we need the spin connection associated to the Levi-Civita connection, which is given by
\begin{equation}
	\widetilde{\omega}^A{}_B = -\frac{q}{2} \epsilon^A{}_{BC}\theta^C
	\label{eq:c32-23}
\end{equation} 
or, on the worldline,
\begin{equation}
	\widetilde{\omega}_\tau{}^A{}_B = -\frac{q}{2} \epsilon^A{}_{BC}E^C.
	\label{eq:c32-24}
\end{equation}
Then, using also the frame equation (\ref{eq:c32-6}) for $\dot{u}^A$,
\begin{align}
\widetilde{\nabla}_\tau u^A &=  \dot{u}^A + \widetilde{\omega}_\tau{}^A{}_B u^B =
-(\frac{m}{s}+\frac{q}{2}) \epsilon^A{}_{BC} u^B E^C.
\label{eq:c32-25}
\end{align}
Notice that this equation is valid both in the regular and critical sectors, but it is trivial in the regular one, for which $\epsilon^A{}_{BC} u^B E^C=0$. In the critical sector, it boils down to
\begin{align}
	\widetilde{\nabla}_\tau u^A &=  
	-\frac{m}{2s} \epsilon^A{}_{BC} u^B E^C,
	\label{eq:c32-26}
\end{align}
and then one can immediately get the Levi-Civita covariant transport for $S^{AB}$,
\begin{equation}
\widetilde{\nabla}_\tau S^{AB} = m (u^B E^A-   u^A E^B) .
\label{eq:c32-27}
\end{equation}
It remains to show that (\ref{eq:c32-21}) can be rewritten in Papapetrou form,  in terms of $S^{AB}$ and the curvature tensor $\widetilde{R}_{MNAB}$,  
\begin{equation}
  \widetilde{\nabla}_\tau p_M = -\frac{1}{2}\widetilde{R}_{MNAB} \dot{x}^N S^{AB}.
\label{eq:c32-28}
\end{equation}
Using the constant (sectional) curvature identity
\begin{equation}
	\widetilde{R}_{MNAB} = -\frac{q^2}{4} (\theta_{AM}\theta_{BN}-\theta_{AN}\theta_{BM})
\end{equation}
and the definition of $S^{AB}$ one immediately obtains 
\begin{equation}
	-\frac{1}{2}\widetilde{R}_{MNAB} \dot{x}^N S^{AB} = -\frac{q^2 s}{2} \epsilon_{CAB} u^C E^B \theta^A{}_M.
\end{equation}
This becomes the forcing term of (\ref{eq:c32-21}) once $\dot{u}^A=q\epsilon^A{}_{BC} u^B E^C$ is used and proves (\ref{eq:c32-28}).

The Papapetrou interpretation is therefore sector-dependent. In the
regular sector, the constraint $\epsilon^A{}_{BC}u^BE^C=0$ forces
$E^A$ to be parallel to $u^A$. Consequently the spin-curvature force
vanishes, the momentum is proportional to the metric velocity, and the
Levi-Civita equation reduces to the ordinary geodesic equation after an
affine reparametrisation.

In the critical sector, $m+qs=0$, the transverse part of $E^A$ is not
constrained to vanish. Equivalently, the momentum--velocity map is
degenerate: $p_M=-m u_A\theta^A{}_M$ fixes $u^A$, but not the full
velocity $E^A=\theta^A{}_M\dot x^M$. The Papapetrou equation is then a
first-order momentum--spin transport equation, but it does not determine
a unique spacetime trajectory without fixing the additional gauge
freedom.

\subsection{Canonical structure of the massive particle}
\label{sec:massive-canonical}

In this section we analyse the canonical structure of the teleparallel
relativistic spinning particle. From the lagrangian
\begin{equation}
	L=-m\,u_A \theta^A_M(x)\dot{x}^M	+	s(\cosh v-1)\dot\varphi,
	\label{eq:rel-spin-lagrangian-canonical-analysis}
\end{equation}
we can obtain the  canonical momenta conjugate to \(x^M,v,\varphi\)
\begin{align}
	p_M&=\frac{\partial L}{\partial \dot x^M}
	=
	-m\,u_A \theta^A{}_M(x),
	\label{eq:pM-primary-rel}\\
	\pi_v&=\frac{\partial L}{\partial \dot v}=0,
		\label{eq:boost-momenta-primary-rel1}\\
	\pi_\varphi&=\frac{\partial L}{\partial \dot\varphi}=
	s(\cosh v-1).
	\label{eq:boost-momenta-primary-rel2}
\end{align}

Introducing frame components
\begin{equation}
	p_A=e_A{}^M(x)p_M,
	\label{eq:frame-momentum-rel}
\end{equation}
we obtain the primary constraints
\begin{equation}
	\Phi_A:=p_A+mu_A\approx0,
	\label{eq:PhiA-rel}
\end{equation}
together with
\begin{equation}
	\Psi_v:=\pi_v\approx0,
	\qquad
	\Psi_\varphi:=\pi_\varphi-s(\cosh v-1)\approx0 .
	\label{eq:Psi-rel}
\end{equation}
The canonical hamiltonian vanishes, $H_c=0$, as expected for a reparametrisation-invariant first-order action.

Since \(u_Au^A=-1\), the three constraints \(\Phi_A\) imply the
mass-shell constraint
\begin{equation}
	C:=(p^A-m u^A)\Phi_A=p_Ap^A+m^2
	=
	g^{MN}(x)p_Mp_N+m^2
	\approx0 .
	\label{eq:mass-shell-rel}
\end{equation}

It is useful to separate the two remaining components of \(\Phi_A\) by
projecting along the tangent directions to the unit hyperboloid:
\begin{align}
	\chi_v:&=\partial_vu^A\Phi_A = \partial_v u^A p_A,
	\label{eq:chi-rel1}\\
	\chi_\varphi:&=\partial_\varphi u^A\Phi_A = \partial_\varphi u^A p_A,
\end{align}
where $u^A \partial_v u_A = u^A \partial_\varphi u_A=0$ have been used.
Hence,  
$
\{	C, \chi_v, \chi_\varphi\}
$
is equivalent to $\{\Phi_A\}_{A=0,1,2}$, away from the coordinate singularities of
the parametrisation \(v,\varphi)\).

The total hamiltonian may be written as
\begin{equation}
	H_T
	=
	\frac{\lambda}{2}C
	+\alpha^v\chi_v+\alpha^\varphi\chi_\varphi
	+\beta^v\Psi_v+\beta^\varphi\Psi_\varphi .
	\label{eq:HT-rel}
\end{equation}
The elementary Poisson brackets are
\begin{equation}
	\{x^M,p_N\}=\delta^M_N,
	\qquad
	\{v,\pi_v\}=1,
	\qquad
	\{\varphi,\pi_\varphi\}=1 .
	\label{eq:elementary-PB-rel}
\end{equation}
The frame
momenta $p_A$ obey
\begin{align}
	\{p_A,p_B\}
	&= \{e_A^M p_M,e_B^N p_N\} = (\partial_M e_A{}^N e_B{}^M - \partial_M e_B{}^N e_A{}^M )p_N.
	\label{eq:frame-PB-rel1}
\end{align}
The right hand-side can be computed from $e_A{}^N \theta^A{}_M = \delta^N{}_M$ and (\ref{eq:c32-9}), and one immediately gets
\begin{align}
	\{p_A,p_B\}
	&= -q \epsilon^C{}_{AB}e^N{}_C p_N = - q \epsilon^C{}_{AB}p_C.
	\label{eq:frame-PB-rel}
\end{align}

Using the explicit parametrisation of $u^A$ in terms of $v,\varphi)$,
\begin{equation}
	u^A = (-\cosh v, \sinh v \cos\varphi, \sinh v \sin\varphi),
\end{equation}
and ordering the four constraints as
$
	\Theta_I=(\chi_v,\chi_\varphi,\Psi_v,\Psi_\varphi)
$,
one can compute the matrix of Poisson brackets
\begin{equation}
	M_{IJ}:=\{\Theta_I,\Theta_J\}
	=
	\begin{pmatrix}
		0 & -mq\sinh v & m & 0\\
		mq\sinh v & 0 & 0 & m\sinh^2 v\\
		-m & 0 & 0 & s\sinh v\\
		0 & -m\sinh^2 v & -s\sinh v & 0
	\end{pmatrix}.
	\label{eq:M-matrix-rel}
\end{equation}
Its determinant is
\begin{equation}
	\det M
	=
	m^2\sinh^4 v\,(m+qs)^2 .
	\label{eq:detM-rel}
\end{equation}
Therefore, in the regular dynamical sector $m+qs\neq 0$ the 4 constraints $\chi_v,\chi_\varphi,\Psi_v,\Psi_\varphi$ are second class, and one can check that the remaining constraint 
$C=p_Ap^A+m^2$ is first class.

Computing the inverse of $M$ and defining
\begin{equation}
	V_v^M:=e_A{}^M\partial_vu^A,
	\qquad
	V_\varphi^M:=e_A{}^M\partial_\varphi u^A,
	\label{eq:V-fields-rel}
\end{equation}
the relevant Dirac brackets $\{F,G\}_D
=
\{F,G\}
-
\{F,\Theta_I\}(M^{-1})^{IJ}\{\Theta_J,G\}$ can be written as
\begin{equation}
	\{v,\varphi\}_D
	=
	-\frac{q}{(m+qs)\sinh v},
	\label{eq:vphi-DB-rel}
\end{equation}
\begin{equation}
	\{x^M,v\}_D
	=
	-\frac{1}{m+qs}\,V_v^M,
	\label{eq:xv-DB-rel}
\end{equation}
\begin{equation}
	\{x^M,\varphi\}_D
	=
	-\frac{1}{(m+qs)\sinh^2 v}\,V_\varphi^M,
	\label{eq:xphi-DB-rel}
\end{equation}
and
\begin{align}
	\{x^M,x^N\}_D
	&=
	-\frac{s}{m(m+qs)\sinh v}
	\left(
	V_\varphi^M V_v^N
	-
	V_v^M V_\varphi^N
	\right)\nonumber\\
	&=
	-\frac{s}{m(m+qs)}
	\epsilon^{ABC}u_C\,e_A{}^M e_B{}^N .
	\label{eq:xx-DB-covariant-rel}
\end{align}
In the torsionless limit \(q=0\), this becomes the standard anyonic
noncommutativity of the relativistic spinning particle in \(2+1\)
dimensions \cite{Pryce:1948,Chou:1993,Jackiw:1990ka,Horvathy:2002vt}.  

Going back to $H_T$, preservation of the second-class constraints
fixes the arbitrary functions $\alpha^v$, $\alpha^\varphi$, $\beta^v$ and
$\beta^\varphi$, while $\lambda$ remains arbitrary and is associated with
the existence of gauge transformations, corresponding to worldline
reparametrisations, generated by the first-class constraint through
the generator $\varepsilon\,\frac{C}{2}$.

The physical phase-space dimension in the regular branch is
$10-4-2\times 1=4$,
where four dimensions are removed by the four second-class constraints and
two by the first-class constraint \(C\). 
After fixing the reparametrisation gauge, the 4 remaining phase-space physical degrees of freedom can be chosen as $x^a, p_a$, with $a=1,2$. The spin does not add an independent propagating degree of freedom; instead
it deforms the reduced symplectic structure. Details of the gauge fixing procedure are provided in Appendix \ref{sec:gf}.

In the critical sector $m=-qs$
the matrix \(M\)  has rank two.  Hence two of the four constraints
\(\Theta_I\) become first class.  A convenient basis is
\begin{equation}
	\Gamma_v
	=
	\Psi_v+\frac{1}{q\sinh v}\chi_\varphi,
	\label{eq:Gammav-rel}
\end{equation}
\begin{equation}
	\Gamma_\varphi
	=
	\Psi_\varphi-\frac{\sinh v}{q}\chi_v,
	\label{eq:Gammaphi-rel}
\end{equation}
valid for \(q\neq0\) and away from \(\sinh v=0\).  One may take
$\chi_v$, $\chi_\varphi$
as the remaining second-class pair.  Their bracket is
\begin{equation}
	\{\chi_v,\chi_\varphi\}=-mq\sinh v .
	\label{eq:critical-second-class-bracket-rel}
\end{equation}
The first-class set is $C$, $\Gamma_v$, $\Gamma_\varphi$
and the total hamiltonian for the critical sector may be written as
\begin{equation}
	H_T
	=
	\frac{\lambda}{2}C
	+
	\rho^v\Gamma_v+\rho^\varphi\Gamma_\varphi
	+
	\sigma^v\chi_v+\sigma^\varphi\chi_\varphi .
	\label{eq:HT-critical-rel}
\end{equation}
Preservation of the second-class constraints now fixes only
\(\sigma^v,\sigma^\varphi\), while \(\lambda,\rho^v,\rho^\varphi\) remain
arbitrary.

The two new gauge transformations act on the boost variables as
\begin{equation}
	\delta v=\rho^v,
	\qquad
	\delta\varphi=\rho^\varphi ,
	\label{eq:critical-gauge-vphi-rel}
\end{equation}
while for  the spacetime coordinates,
\begin{equation}
	\delta x^M
	=
	\frac{\rho^v}{q\sinh v}\,V_\varphi^M
	-\frac{\rho^\varphi\sinh v}{q}\,V_v^M.
	\label{eq:critical-gauge-x-rel}
\end{equation}
Thus, at \(m=-qs\), two directions which were physical in the regular
symplectic form become gauge directions.

The physical phase-space dimension in the critical sector is
$10-2 - 2\times3 = 2$, where now 2 degrees of freedom are removed by
the 2 second class constraints and the other 6 by the 3 first class
ones.  This is consistent with the drop in dimension of the
corresponding coadjoint orbit at the exceptional locus.\footnote{We
  should remind the reader that although the dimensions of the
  coadjoint orbit and the (reduced) phase space coincide, it is
  important to remark that they are not the same space: a particle
  trajectory is a point in the coadjoint orbit (the momenta of the
  particle), whereas it is the integral curve of a hamiltonian vector
  field in the phase space (the actual trajectory).}

The extra gauge transformations can be used to set the Goldstone
degrees of freedom to constant values and, together with the fixing of
the reparametrisation gauge transformation, this leaves $x^a$,
$a=1,2$, as the remaining physical degrees of freedom, with nontrivial
Dirac brackets and evolving with the gauge fixed hamiltonian
$H_{red}=-p_0$, with $p_0$ expressed in terms of $x^a$. As in the
regular case, details have been deferred to Appendix~\ref{sec:gf}.

\section{Massless particle dynamics in Mielke--Baekler spacetimes}
\label{sec:massless-particles}

We now consider massless particles propagating in a Mielke--Baekler
spacetime.

\subsection{Massless particle orbits}
\label{sec:massl-part-orbits}

The momentum of such a particle is null and, without loss of
generality, we can take it to be proportional to $\pi^0 - \pi^1$.
Using the formulae \eqref{eq:coadjoint-action}, we see that the
stabiliser subalgebra of this momentum is abelian and two-dimensional,
spanned by $J_0 + J_1$ and $P_0 + P_1$.  The most general momentum
stabilised by this subalgebra is of the form
\begin{equation}
	\label{eq:massless-moment}
	\alpha = \mu (\pi^0 - \pi^1) - s (\lambda^0 - \lambda^1).
\end{equation}
Although the generic stabiliser subalgebra of such a momentum is the
two-dimensional subalgebra above, for special values of $\mu$ and
$s$, it can be larger.

Let us write $X = a^A J_A + b^A P_A \in \g$ and consider the coadjoint
action of $X$ on the above $\alpha$.  A calculation using
equations~\eqref{eq:coadjoint-action}, yields
\begin{multline}
	\ad_X^* \alpha = (-s a^2 + \mu b^2) (\lambda^0 - \lambda^1) +
	\left((-p s + q \mu) b^2 + \mu a^2\right) (\pi^0 - \pi^1)\\ + (-s (a^1 -
	a^0) + \mu (b^1 - b^0)) \lambda^2 + \left(\mu (a^1 - a^0) + (-ps + q
	\mu)(b^1-b^0)\right) \pi^2.
\end{multline}
Therefore $X \in \g_\alpha$ if $a^A,b^A$ satisfy the following system of linear homogeneous equations
\begin{equation}
	\begin{pmatrix}
		-s & \mu & 0 & 0 \\
		\mu & -p s + q \mu & 0 & 0\\
		0 & 0 & -s & \mu \\
		0 & 0 & \mu & -p s + q \mu
	\end{pmatrix}
	\begin{pmatrix}
		a^2\\
		b^2\\
		a^1 - a^0\\
		b^1 - b^0
	\end{pmatrix} =
	\begin{pmatrix}
		0\\
		0\\
		0\\
		0
	\end{pmatrix}.
\end{equation}
The determinant of the matrix is $(p s^2 -q s \mu -
\mu^2)^2$.  For generic values of $s,\mu$ this is nonzero and
hence the stabiliser subalgebra is spanned by $J_0 + J_1$ and $P_0 +
P_1$, so that the coadjoint orbit $\eO_\alpha$ has dimension $4$.
However if $\mu^2 = s (ps - q \mu)$ (and $\alpha \neq 0$),
the stabiliser subalgebra is four-dimensional and spanned by $J_0 + J_1$,
$P_0 + P_1$, $\mu J_2 +s P_2$ and $\mu (J_0 - J_1) + s(P_0
- P_1)$.  In this critical case, the orbit $\eO_\alpha$ has dimension
$2$. As in the massive case, this fact will resurface in the analysis of the massless  particle action.

We introduce the light-cone coordinates
\begin{equation}
	x^\pm
	=
	\frac{1}{\sqrt{2}}
	\left(x^0\pm x^1\right),
\end{equation}
in terms of which the light-cone line element is given by
\begin{equation}
	\dd s^2=-2 \dd x^+ \dd x^-+(\dd x^2)^2=\eta_{AB}\dd x^A \dd x^B.
\end{equation}
In the ordered light-cone basis \((+,2,-)\), the flat metric is
\begin{equation}
	\eta_{AB}
	=
	\begin{pmatrix}
		0 & 0 & -1 \\
		0 & 1 & 0 \\
		-1 & 0 & 0
	\end{pmatrix}.
	\label{lcmetric}
\end{equation}
We will also use the Lie algebra generator combinations
\begin{equation}
	P_\pm
	=
	\frac{1}{\sqrt{2}}
	\left(P_0\pm P_1\right),
	\qquad
	J_\pm
	=
	\frac{1}{\sqrt{2}}
	\left(J_0\pm J_1\right).
\end{equation}

\subsection{Homogeneous space construction}
\label{sect3}

We start with  a preliminary analysis of the homogeneous MB space,
defined as the space of cosets $\mathrm{MB}|_{p=0}/\mathrm{Lorentz}$,
where $\mathrm{MB}|_{p=0}$ is the Lie group with Lie algebra
\eqref{eq:MB-algebra} with $p=0$ and where $\mathrm{Lorentz}$ denotes
the subgroup with Lie algebra spanned by $\{J_+,J_2,J_-\}$.  These
results will be useful for the construction of the MB massless
particle action.

A local coset representative is given by
\begin{equation}
  g_0(x)=\exp(x^AP_A)=\exp(x^+P_++x^-P_-+x^2P_2).
\end{equation}
Notice that this is the same construction presented for the massive
case, but with a different basis, and hence the homogeneous space is
exactly the same but in different coordinates.  If, as in the massive
framework, $\Omega_0= g_0^{-1}\dd g_0=\theta^A P_A$, $A=+,-,2$, it is
easy to see that
\begin{align}
\theta^+ ={}&
	\left[
	1-Ax^2
	+
	B\left(
	-x^+x^-+(x^2)^2
	\right)
	\right]\dd x^+
	+
	B(x^+)^2\dd x^-
	+
	x^+\left(A-Bx^2\right)\dd x^2,
	\\[1mm]
	\theta^-
	={}&
	B(x^-)^2\dd x^+
	+
	\left[
	1+Ax^2
	+
	B\left(
	-x^+x^-+(x^2)^2
	\right)
	\right]\dd x^-
	-
	x^-\left(A+Bx^2\right)\dd x^2,
	\\[1mm]
	\theta^2
	={}&
	-x^-\left(A-Bx^2\right)\dd x^+
	+
	x^+\left(A+Bx^2\right)\dd x^-
	+
	\left[
	1-2Bx^+x^-
	\right]\dd x^2,
	\label{mcE}
\end{align}
where
\begin{equation}
	A(\rho)=\frac{\cosh(q\rho)-1}{q\rho^2},
	\qquad
	B(\rho)=\frac{\sinh(q\rho)-q\rho}{q\rho^3}.
\end{equation}

\subsection{The MB massless particle}
\label{sec:massless}

To build the particle action, we need to use the little group of the momentum reference frame of the considered particle. For a massless particle  the little group is $E(1)$, generated, in light-cone coordinates, by \(J_+\), while the broken Lorentz generators are \(J_-\) and \(J_2\), although one could also exchange the roles of $J_+$ and $J_-$.
A convenient coset representative is then
\begin{equation}
	g(x,u,\varphi)=g_0(x)b(u,\varphi),
	\qquad
	b(u,\varphi)=\exp(uJ_-)\exp(\varphi J_2),
\end{equation}
and the associated Maurer-Cartan form is
\begin{equation}
	\Omega= g^{-1}\dd g=b^{-1}
	\Omega_0 b+b^{-1} db= \vartheta^A P_A+\varpi^A J_A .
\end{equation}
The dressed translational forms are
\begin{equation}
	\begin{pmatrix}
		\vartheta^+ \\
		\vartheta^2 \\
		\vartheta^-
	\end{pmatrix}
	=
	\Lambda(u,\varphi)
	\begin{pmatrix}
		\theta^+ \\
		\theta^2 \\
		\theta^-
	\end{pmatrix},
\end{equation}
with the Lorentz transformation, in the $+2-$ ordered light-cone basis,
\begin{equation}
	\Lambda(u,\varphi)
	=
	\begin{pmatrix}
		e^{-\varphi} & 0 & 0 \\[1mm]
		-u & 1 & 0 \\[1mm]
		\frac12 e^\varphi u^2 & -e^\varphi u & e^\varphi
	\end{pmatrix},\quad \Lambda^T\eta\,\Lambda=\eta,
\end{equation}
while the Lorentz forms are
\begin{equation}
	\varpi^-=e^\varphi du,
	\qquad
	\varpi^2=d\varphi,
	\qquad
	\varpi^+=0 .
\end{equation}

As in the massive case, we reserve the notation \(E^A\) for the
worldline pullback components of the spacetime coframe:
\begin{equation}
	\gamma^*\theta^A = E^A d\tau,
	\qquad
	E^A := \theta^A{}_M(x)\dot x^M .
\end{equation}
Similarly, for the dressed translational forms we write
\begin{equation}
	\gamma^*\vartheta^A = \mathcal E^A d\tau,
	\qquad
	\mathcal E^A := \Lambda^A{}_B(u,\varphi)E^B .
\end{equation}
In particular,
\begin{equation}
	\mathcal E^-
	=
	e^\varphi
	\left(
	E^- - uE^2 + \frac12 u^2 E^+
	\right).
\end{equation}
Following the discussion leading to (\ref{eq:massless-moment}), we are led to consider the two-parameter massless action
\begin{equation}
	S_{\mu,s}
	=
	\int
	\left(
	-\mu\,\vartheta^- + s\,\varpi^-
	\right).
\end{equation}
After pullback to the worldline, the lagrangian is
\begin{equation}
	L
	=
	-\mu\,\mathcal E^-
	+
	se^\varphi \dot u .
	\label{eq:massless-two-param-lagrangian}
\end{equation}
Introducing
\begin{equation}
	N_A(u,\varphi)=e^\varphi n_A(u),
	\qquad
	n_A(u)=
	\left(
	\frac12u^2,-u,1
	\right),
\end{equation}
we may equivalently write
\begin{equation}
	\mathcal E^- = N_AE^A,
\end{equation}
and hence
\begin{equation}
	L
	=
	-\mu N_AE^A
	+
	se^\varphi\dot u
	=
	-\mu N_A\theta^A{}_M(x)\dot x^M
	+
	se^\varphi\dot u .
\end{equation}

\subsection{Canonical analysis}
\label{sec:canonical-analysis}

From (\ref{eq:massless-two-param-lagrangian}),	the canonical momenta conjugate to \(x^M,u,\varphi\) are
\begin{align}
	p_M
	&:=
	\frac{\partial L}{\partial \dot x^M}
	=
	-\mu N_A\theta^A{}_M ,
	\label{eq:massless-pM}
	\\
	p_u
	&:=
	\frac{\partial L}{\partial \dot u}
	=
	s e^\varphi ,
	\label{eq:massless-pu}
	\\
	p_\varphi
	&:=
	\frac{\partial L}{\partial \dot\varphi}
	=
	0 .
	\label{eq:massless-pphi}
\end{align}
As in the massive case, using the inverse frame \(e_A{}^M\)  to define  the frame components of the momentum, $p_A=e_A{}^M p_M$,
the primary constraints can be written as
\begin{align}
	\Phi_A
	&:=
	p_A+\mu N_A
	\approx 0 ,
	\label{eq:massless-PhiA}
	\\
	\Psi_u
	&:=
	p_u-se^\varphi
	\approx 0 ,
	\label{eq:massless-Psiu}
	\\
	\Psi_\varphi
	&:=
	p_\varphi
	\approx 0 .
	\label{eq:massless-Psiphi}
\end{align}
The canonical hamiltonian vanishes, $H_c=0$, 
as expected for a first-order reparametrisation-invariant action.

As in the massive case, the frame momenta obey the Poisson algebra
\begin{equation}
	\{p_A,p_B\}
	=
	-q\epsilon^C{}_{AB}p_C ,
	\label{eq:massless-frame-momentum-bracket}
\end{equation}
since this depends only  on the Maurer--Cartan equation for the teleparallel coframe.
The remaining canonical Poisson brackets are
\begin{equation}
	\{x^M,p_N\}=\delta^M{}_N,
	\qquad
	\{u,p_u\}=1,
	\qquad
	\{\varphi,p_\varphi\}=1,
	\label{eq:massless-canonical-PB}
\end{equation}
and one also has
\begin{equation}
	\{x^M,p_A\}=e_A{}^M(x).
	\label{eq:massless-x-frame-momentum-PB}
\end{equation}

Since \(N_A\) is null, the constraints \(\Phi_A\approx0\) imply the mass-shell
constraint
\begin{equation}
	C:=(p^A-\mu N^A)\Phi_A = p_Ap^A
	=
	G^{MN}(x)p_Mp_N
	\approx 0 ,
	\label{eq:massless-C}
\end{equation}
where
\begin{equation}
  G^{MN}=e_A{}^M e_B{}^N\eta^{AB}
  \label{eq:massless-inverse-metric}
\end{equation}
is the inverse of the metric $G_{MN}$ in (\ref{eq:gMB}).
The quantity \(C=p_Ap^A\) is a Casimir of the algebra
\eqref{eq:massless-frame-momentum-bracket}, and has trivially zero Poisson brackets with all the internal variables $u$, $p_u$, $\varphi$, $p_\varphi$;  hence it is first class.

To display the remaining constraints, introduce the null triad
\begin{equation}
	n_A
	=
	\left(
	\frac12u^2,\,-u,\,1
	\right),
	\qquad
	m_A
	=
	\partial_u n_A
	=
	(u,-1,0),
	\qquad
	\ell_A=(1,0,0).
	\label{eq:massless-null-triad}
\end{equation}
It obeys
\begin{equation}
	n^2=\ell^2=0,
	\qquad
	m^2=1,
	\qquad
	n\cdot\ell=-1,
	\qquad
	n\cdot m=\ell\cdot m=0 .
	\label{eq:massless-null-triad-products}
\end{equation}
A convenient basis for the two components of \(\Phi_A\) transverse to the
mass-shell constraint is
\begin{align}
	R
	&:=
	m^A\Phi_A
	=
	m^A p_A
	\approx 0 ,
	\label{eq:massless-R}
	\\
	S
	&:=
	\ell^A\Phi_A
	=
	\ell^A p_A-\mu e^\varphi
	\approx 0 .
	\label{eq:massless-S}
\end{align}
Together, the set $(C,R,S)$
is equivalent to the three constraints \(\Phi_A\). We now collect the four constraints
$\Theta_I=(R,S,\Psi_u,\Psi_\varphi)$, and compute their matrix of Poisson brackets
\begin{equation}
	M_{IJ}
	:=
	\{\Theta_I,\Theta_J\}
	\approx
	e^\varphi
	\begin{pmatrix}
		0 & -\mu q & \mu & 0\\
		\mu q & 0 & 0 & -\mu\\
		-\mu & 0 & 0 & -s\\
		0 & \mu & s & 0
	\end{pmatrix},
	\label{eq:massless-MIJ}
\end{equation}
where $\approx$ stands for a weak equality, using the primary constraints $p_A\approx -\mu e^\varphi n_A$.
The determinant is
\begin{equation}
	\det M
	=
	e^{4\varphi}\mu^2(\mu+qs)^2 .
	\label{eq:massless-detM}
\end{equation}
This is the hamiltonian manifestation of the sector structure for the massless particle.

In the regular sector, $\mu\neq -qs$,
the four constraints \(\Theta_I\) are second class.  The only first-class
constraint is the mass-shell constraint \(C\approx0\), and the  total hamiltonian is
\begin{equation}
	H_T
	=
	\frac{\lambda}{2}C+\lambda^I\Theta_I .
	\label{eq:massless-HT-regular}
\end{equation}
Preservation of the second-class constraints fixes the four multipliers
\(\lambda^I\), while \(\lambda\) remains arbitrary and generates worldline
reparametrisations.

Computing the inverse of the matrix \eqref{eq:massless-MIJ}
allows one to get the  	 Dirac brackets
\begin{align}
	\{u,\varphi\}_D
	&=
	\frac{q e^{-\varphi}}{\mu+qs},
	\label{eq:massless-u-phi-DB}
	\\
	\{x^M,u\}_D
	&=
	-\frac{e^{-\varphi}}{\mu+qs}
	m^Ae_A{}^M,
	\label{eq:massless-x-u-DB}
	\\
	\{x^M,\varphi\}_D
	&=
	\frac{e^{-\varphi}}{\mu+q s}
	\ell^Ae_A{}^M .
	\label{eq:massless-x-phi-DB}
\end{align}
The spacetime coordinates have the Dirac bracket
\begin{equation}
	\{x^M,x^N\}_D
	=
	\frac{s e^{-\varphi}}{\mu(\mu+qs)}
	\left(
	m^Ae_A{}^M\ell^Be_B{}^N
	-
	\ell^Ae_A{}^M m^Be_B{}^N
	\right).
	\label{eq:massless-x-x-DB}
\end{equation}
Thus the Lorentz-orbit parameter \(s\) produces a non-trivial coordinate
Dirac bracket.  For \(s=0\), this bracket vanishes and one recovers the
usual spinless massless particle.

After imposing the second-class constraints, the dynamics is generated by
\begin{equation}
	H_{\rm red}
	=
	\frac{\lambda}{2}C .
	\label{eq:massless-Hred-regular}
\end{equation}
Since \(C\) has vanishing brackets with the second-class constraints, its
hamiltonian flow is equivalently computed with the ordinary Poisson bracket.
Hence
\begin{align}
	\dot x^M
	&=
	\lambda G^{MN}(x)p_N,
	\label{eq:massless-xdot-regular}
	\\
	\dot p_M
	&=
	-\frac{\lambda}{2}
	\partial_MG^{NP}(x)p_Np_P .
	\label{eq:massless-pdot-regular}
\end{align}
These are the usual Hamilton equations for a null geodesic in the MB/AdS
metric, with \(\lambda\) playing the role of the worldline einbein.

As in the massive case, the physical phase-space dimension  is $10-4-2\times 1=4$.

In the critical sector, $\mu=-qs$,
the matrix \(M_{IJ}\) has rank two.  Two of the four constraints
\(\Theta_I\) become first class.  A convenient first-class basis is
\begin{align}
	\Gamma_u
	&:=
	\Psi_u+\frac{1}{q}S,
	\label{eq:massless-Gammau}
	\\
	\Gamma_\varphi
	&:=
	\Psi_\varphi+\frac{1}{q}R.
	\label{eq:massless-Gammaphi}
\end{align}
Together with \(C\), these form a first-class set.
The remaining two constraints may be chosen to be \(R\) and \(S\), with Poisson bracket
$\{R,S\}
=
-\mu q e^\varphi
$,
which is non-zero as long as \(\mu q \neq0\).

The critical total hamiltonian may be written as
\begin{equation}
	H_T
	=
	\frac{\lambda}{2}C
	+
	\rho_u\Gamma_u
	+
	\rho_\varphi\Gamma_\varphi
	+
	\alpha R+\beta S .
	\label{eq:massless-HT-critical}
\end{equation}
Preservation of \(R\) and \(S\) fixes \(\alpha\) and \(\beta\), while
\(\lambda,\rho_u,\rho_\varphi\) remain arbitrary.  The arbitrariness of
\(\rho_u\) and \(\rho_\varphi\) is the hamiltonian manifestation of the two
extra gauge symmetries that appear at the critical locus.

The additional gauge transformations generated by \(\rho_u\Gamma_u+\rho_\varphi\Gamma_\varphi\) are
\begin{align}
	\delta u
	&=
	\rho_u ,
	\label{eq:massless-critical-deltau}
	\\
	\delta\varphi
	&=
	\rho_\varphi ,
	\label{eq:massless-critical-deltaphi}
	\\
	\delta x^M
	&=
	\frac{\rho_u}{q}\ell^Ae_A{}^M
	+
	\frac{\rho_\varphi}{q}m^Ae_A{}^M .
	\label{eq:massless-critical-deltax}
\end{align}
Therefore two directions which are physical in the regular symplectic form
become gauge directions at \(\mu=-qs\), and
the physical phase-space dimension is  now $10-2-2\times3=2$.

\subsection{Noether symmetries of the  massless MB particle}
\label{sec:conformal-symmetry}
Consider a point-canonical generator\footnote{One could consider the addition of a boundary term $f(x,u,\varphi)$, but it can be shown that the resulting equations force it to be constant, and hence can be disregarded.} 
\begin{equation}
	\mathcal G
	=\alpha^M(x)p_M+\beta(x,u,\varphi)p_u+\gamma(x,u,\varphi)p_\varphi.
	\label{eq:point-generator}
\end{equation}
It generates
\begin{equation}
	\delta x^M=\alpha^M,
	\qquad
	\delta u=\beta,
	\qquad
	\delta\varphi=\gamma.
\end{equation}
Notice that we are restricting to point transformations of $x$ that depend solely on $x$; thus they are not the most general point transformations, but are sufficient to include conformal transformations.

We want the above $ \mathcal G$ to generate Noether transformations. In the present setup, with identically zero canonical hamiltonian, this is equivalent to
\begin{equation}
	\delta\Phi_A := \{\Phi_A, \mathcal G\}\simeq 0,\quad 
	\delta\Psi_u := \{\Psi_u, \mathcal G\}\simeq 0,\quad
	\delta\Psi_\varphi := \{\Psi_\varphi, \mathcal G\}\simeq 0,
	\label{eq:Killing1}
\end{equation}
with $\simeq$ meaning equality up to primary constraints.

Since the Lie derivative of a form is a form of the same degree and $\{\theta^A\}$ is a co-frame basis, we can define $K^A{}_B$ by
\begin{equation}
	\Lie_\alpha\theta^A=K^A{}_B\theta^B.
\end{equation}
Then, using $e_A(\theta^B)=\delta_A{}^B$,
one has 
\begin{equation}
	\Lie_\alpha e_A=-K^B{}_Ae_B,
\end{equation}
and also, since this is just the Lie derivative of the vector field $e_A$ along the vector field $\alpha=\alpha^M\partial_M$,
\begin{equation}
	-K^B{}_Ae_B{}^M = \alpha^N\partial_Ne_A{}^M - e_A{}^N\partial_N\alpha^M.
	\label{eq:Kalpha}
\end{equation}

From this, using $e_A{}^M\theta^B{}_M=\delta_A{}^B$, one can obtain an explicit expression for $K^A{}_B$,
\begin{equation}
	K^A{}_B = - \theta^A{}_M \left(\alpha^N \partial_N e_B{}^M- e_B{}^N \partial_N\alpha^M \right).
	\label{eq:KAB1}
\end{equation}
An alternative expression, in terms of the derivatives of $\theta^A{}_M$ instead of those of $e_A{}^M$, is given by
\begin{equation}
	K^A{}_B = e_B{}^M \left(\alpha^N \partial_N \theta^A{}_M + \theta^A{}_N \partial_M\alpha^N \right), \label{eq:KAB2}
\end{equation}
and still another useful form is, in terms of $\alpha^A:= \imath_\alpha\theta^A$,
\begin{equation}
	K^A{}_B 
	= e_B\alpha^A + q \epsilon^A{}_{BC} \alpha^C,
	\label{eq:KAB3}
\end{equation}
where the Maurer-Cartan equation (\ref{eq:MC}) has been used to evaluate $\imath_\alpha\dd\theta^A$. Lowering indices we have
\begin{equation}
	K_{AB} 
	= e_B\alpha_A + q \epsilon_{ABC} \alpha^C.
	\label{eq:KAB4}
\end{equation}
Only the symmetric part of $K$
contributes to the variation of the space-time metric $G_{MN}=\eta_{AB} \theta^A{}_M\theta^B{}_N$ under the field $\alpha$,
\begin{equation}
	{\cal L}_\alpha G_{MN}= (K_{AB}+K_{BA}) \theta^A{}_M \theta^B{}_N = 2 K_{(AB)} \theta^A{}_M \theta^B{}_N.
	\label{eq:LalphaG}
\end{equation}

Evaluating the equations in (\ref{eq:Killing1}) one obtains the Noether symmetry equations (NSE)

\begin{align}
	K^B{}_A n_B+\beta m_A+\gamma n_A
	-\frac{s}{\mu}e_A\beta&=0,
	\label{eq:noether-cond-1}\\
	s(\partial_u\beta+\gamma)&=0,
	\label{eq:noether-cond-2}\\
	s \partial_\varphi\beta&=0.
	\label{eq:noether-cond-3}
\end{align}

Notice that, in general, the right-hand side of
equation~\eqref{eq:LalphaG} is neither zero nor proportional to
$G_{MN}$, and a general $\alpha$ is thus neither a Killing nor a
conformal Killing vector field.  We will see, however, that the NSE do
actually force $\alpha$ to be (conformal) Killing, depending on
whether $s$ is zero or not.

For $s=0$ only (\ref{eq:noether-cond-1}) survives, and contracting the first equation with $n^A$ one immediately gets
\begin{equation}
	n^A K_{BA}n^B =0.
	\label{eq:noether-4}
\end{equation}
We now decompose the symmetric part $K_{(AB)}$ of $K_{AB}$ as
$K_{(AB)} = \Theta \eta_{AB} + \Sigma_{AB}$,
with $\Sigma_{AB}$ traceless, $\eta^{AB}\Sigma_{AB}=0$. Since $n^2=0$, (\ref{eq:noether-4}) implies
$n^A \Sigma_{BA}n^B = 0$.
Using that this must hold for arbitrary $u$ and the fact   that $\Sigma$ is symmetric and traceless one can show that $\Sigma=0$, so that
\begin{equation}
	K_{(AB)} = \Theta(x)\eta_{AB}
	\label{eq:KABs0}
\end{equation}
and 
\begin{equation}
	{\cal L}_\alpha G_{MN}= 2\Theta(x) \eta_{AB}\theta^A{}_M \theta^B{}_N =2 \Theta(x) G_{MN}.
	\label{eq:LalphaGs0}
\end{equation}
Hence, for $s=0$, $\alpha$ is a conformal Killing vector field, with conformal factor $\Theta(x)$.

For $s\neq 0$ the situation is very different. First of all, from (\ref{eq:noether-cond-3}) one has that $\partial_\varphi\beta=0$ and thus $\beta=\beta(x,u)$. Using $\gamma=-\partial_u\beta$ from (\ref{eq:noether-cond-2}) into (\ref{eq:noether-cond-1}) one gets
\begin{equation}
	K^B{}_A n_B + \beta m_A-n_A\partial_u\beta - \frac{s}{\mu}e_A\beta=0.
	\label{eq:noether-cond-4}
\end{equation} 
A little analysis shows that, due to the quadratic terms in $u$ in $n$, the dependence of $\beta(x,u)$ on $u$  must be of the form
\begin{equation}
	\beta(x,u) = \chi_A(x)n^A(u).
	\label{eq:betau}
\end{equation}
Using this form, (\ref{eq:KAB4})  and the identity  (see Appendix \ref{sec:useful-identity})
\begin{equation}
  n_B m_A - n_A m_B=\epsilon_{ABC}n^C,
\label{eq:nontrivial-identity}
\end{equation}
one gets
\begin{equation}
	(e_A\alpha^B+ q \epsilon^B{}_{AC}\alpha^C) n_B - \epsilon_{ABC}\chi^Cn^B - \frac{s}{\mu}n^Be_A\chi_B=0.
	\label{eq:noether-cond-5}
\end{equation}
Since all the terms are proportional to $n^B$, re-arranging terms this leads to
\begin{equation}
	e_A\left(\alpha^B - \frac{s}{\mu} \chi^B\right) - \epsilon^B{}_{CA}\left( \chi^C +q \alpha^C\right)=0.
	\label{eq:noether-cond-6}
\end{equation}
Define now the frame vectors $Y$, $Z$ with components
\begin{equation}
	Y^A := \alpha^A - \frac{s}{\mu} \chi^A,\quad Z^A := q\alpha^A+\chi^A,
	\label{eq:YZ}
\end{equation}
which allow to express (\ref{eq:noether-cond-6}) as
\begin{equation}
	e_A Y_B = \epsilon_{BCA} Z^C.
	\label{eq:YZrel}
\end{equation}
Since the right-hand side is skew-symmetric, we deduce that
\begin{equation}
	e_A Y_B + e_B Y_A =0.
	\label{eq:killingY}
\end{equation}
Equation (\ref{eq:YZrel}) can be inverted for $Z$,
\begin{equation}
	Z_A = \frac{1}{2} \epsilon_A{}^{BC} e_B Y_C.
	\label{eq:ZYrel}
\end{equation}
Using (\ref{eq:ZYrel}), $[e_A,e_B]= q \epsilon^C{}_{AB} e_C$, and (\ref{eq:killingY}) one can see that
\begin{align}
	e_A Z_B &= q \epsilon_{ABC}Z^C,
	 \label{eq:eAZB}
\end{align}
from which
\begin{equation}
	e_A Z_B + e_B Z_A =0.
	\label{eq:killingZ}
\end{equation}
Let us assume now that we are in the regular sector, so that $\mu+qs\neq 0$. Then $\alpha^A$ can be expressed in terms of $Y^A$ and $Z^A$ as
\begin{equation}
	\alpha^A = \frac{1}{\mu+qs}(\mu Y^A+ s Z^A)
\end{equation}
and, in view of (\ref{eq:killingY}) and (\ref{eq:killingZ}),
\begin{equation}
	e_A \alpha_B + e_B \alpha_A =0,
	\label{eq:killingalpha}
\end{equation}
which, according to (\ref{eq:KAB4}), implies that $K_{AB}$ has no symmetric part and thus ${\cal L}_\alpha G=0$. We conclude then that for $s\neq 0$, in the regular sector the field $\alpha$ is a true Killing vector field, not a conformal one. This is in contrast with the $s=0$ case, where a term proportional to $\eta_{AB}$ is allowed for $K_{(AB)}$.

In the critical sector $\mu= -q s$ one cannot express $\alpha$ in terms of the Killing fields $Y$ and $Z$. Instead one has, from the definition of $Y$ in (\ref{eq:YZ}), 
\begin{equation}
	\alpha^A = Y^A+ \Delta^A,\quad \Delta^A := \frac{s}{\mu}\chi^A,
\end{equation}
with $Y^A$ still satisfying (\ref{eq:killingY}) but with $\Delta^A$ non-Killing. Expanding $\Delta^A$ in terms of the null triad $(n,m,\ell)$ one has
\begin{equation}
	\Delta^A = a n^A + b m^A + c \ell^A.
\end{equation}

The most general transformation of $x^M$ is given by the geometric contribution from the field $\alpha^M$ plus the gauge transformations. For the critical sector there are two extra gauge transformations, given by (\ref{eq:massless-critical-deltau})---(\ref{eq:massless-critical-deltax}). Choosing $\rho_u=-q c$, $\rho_\phi=-qb$ we obtain, with $Y^M = e_A{}^{M}Y^A$, $\Delta^M = e_A{}^M\Delta^A$,
\begin{equation}
	\alpha^M + \delta_\text{extra} x^M = Y^M + \Delta^M +  \delta_\text{extra} x^M = Y^M+ a n^A e_A{}^M,
\end{equation}
and two of the non-Killing contributions are removed by the extra gauge transformations. The remaining component can then be compensated using the gauge transformation induced by the mass-shell constraint $C= p_Ap^A= G^{MN}p_M p_N$. Indeed, using the generator $G_\varepsilon = \varepsilon/2 C$ one has
\begin{equation}
	\delta_\varepsilon x^M = \epsilon G^{MN}p_N = \varepsilon p^A e_A{}^M.
\end{equation}
On the constraint surface, $p^A \approx -\mu e^\varphi n^A$. Selecting $\varepsilon=a/(\mu e^{\varphi})$ one has then
\begin{equation}
	\alpha^M + \delta_\text{extra} x^M + \delta_\varepsilon x^M = Y^M,
\end{equation}
which corresponds to a true Killing field. 

In summary, the NSE force the symmetric part of $K_{AB}$ to be either
zero for $s\neq 0$ (in fact, gauge equivalent to zero in the singular
sector) or pure trace for $s=0$. In the first case the corresponding
vector field is a Killing vector of the spacetime metric, while in the
second case it is a conformal Killing vector. Once this restriction
has been taken into account, the Noether symmetry equations can be
solved for $\beta$ and $\gamma$, yielding the transformations of the
internal variables $u$ and $\varphi$.

For the spinless system \(s=0\)  equations (\ref{eq:noether-cond-2})
and (\ref{eq:noether-cond-3}) are empty, while equation
(\ref{eq:noether-cond-1}) becomes, using
$K_{AB}=\Theta\eta_{AB}+\Lambda_{AB}$,  $\Lambda_{AB}=-\Lambda_{BA}$,
\begin{equation}
(\Theta+\gamma)n_A+\beta m_A+\Lambda^B{}_{A}n_B=0.
\label{eq:noether1s0}
\end{equation}
Since \(\Lambda_{AB}\) is antisymmetric,  $n^A\Lambda^B{}_{A}n_B=0$, and
one has   that \(-\Lambda^B{}_{A}n_B\) lies in the two-dimensional space orthogonal to \(n_A\), spanned by \(m_A\) and \(n_A\).  We write
\begin{equation}
-\Lambda^B{}_{A}n_B=\beta_\Lambda m_A+\gamma_\Lambda n_A.
\end{equation}
Contracting  this with \(m^A\)  and $\ell^A$ gives
\begin{align}
	\beta_\Lambda &=-m^A\Lambda^B{}_{A}n_B,\\
	\gamma_\Lambda & =\ell^A\Lambda^B{}_{A}n_B.
\end{align}
Equation (\ref{eq:noether1s0}) is then solved with  $\beta=\beta_\Lambda$ and $\gamma=\gamma_\Lambda-\Theta$, and one finally has
\begin{equation}
	\delta u=-m^A\Lambda^B{}_{A}n_B,
	\qquad
	\delta\phi=\ell^A\Lambda^B{}_{A}n_B-\Theta .
	\label{eq:transuvarphis0}
\end{equation}

Let us now repeat the spinless analysis in the spinning case, for which we have shown that
\begin{equation}
	K_{AB}=\Lambda_{AB},
	\qquad
	\Lambda_{AB}=-\Lambda_{BA}.
\end{equation}
In this case the NSE  become, with  $\lambda:=-\frac{s}{\mu}$,
\begin{align}
	\Lambda^B{}_A n_B+\beta m_A+\gamma n_A
	+\lambda e_A\beta &=0,
	\label{eq:spin-internal-1}
	\\
	\partial_u\beta+\gamma&=0,
	\label{eq:spin-internal-2}
	\\
	\partial_\varphi\beta&=0.
	\label{eq:spin-internal-3}
\end{align}

As in the spinless case, with the same definitions for $\beta_\Lambda$ and $\gamma_\Lambda$,
\begin{equation}
	\Lambda^B{}_A n_B
	=
	-\beta_\Lambda m_A-\gamma_\Lambda n_A .
	\label{eq:spin-Lambda-decomposition}
\end{equation}
Substituting this into \eqref{eq:spin-internal-1} gives
\begin{equation}
	(\beta-\beta_\Lambda)m_A
	+
	(\gamma-\gamma_\Lambda)n_A
	+
	\lambda e_A\beta
	=
	0 .
	\label{eq:spin-deformed-algebraic}
\end{equation}

Projecting \eqref{eq:spin-deformed-algebraic} along $n^A, m^A, \ell^A$,
one obtains
\begin{align}
	n^Ae_A\beta&=0,
	\label{eq:spin-beta-n}
	\\
	\beta+\lambda m^Ae_A\beta&=\beta_\Lambda,
	\label{eq:spin-beta-m}
	\\
	\gamma&=\gamma_\Lambda+\lambda\ell^Ae_A\beta .
	\label{eq:spin-gamma-ell}
\end{align}
Together with \eqref{eq:spin-internal-2}, the last equation can also be written as
\begin{equation}
	\partial_u\beta
	=
	-\gamma_\Lambda-\lambda\ell^Ae_A\beta .
	\label{eq:spin-beta-u}
\end{equation}

Thus the spinning answer is a deformation of the spinless one. Formally, introducing the differential operator
\begin{equation}
	\mathcal D_m:=m^Ae_A,
\end{equation}
equation \eqref{eq:spin-beta-m} gives
\begin{equation}
	\beta
	=
	\left(1+\lambda\mathcal D_m\right)^{-1}\beta_\Lambda ,
	\label{eq:spin-beta-formal}
\end{equation}
with the additional condition
\begin{equation}
	n^Ae_A\beta=0 .
\end{equation}
Then
\begin{equation}
	\gamma
	=
	\gamma_\Lambda+\lambda\ell^Ae_A\beta
	=
	-\partial_u\beta .
	\label{eq:spin-gamma-formal}
\end{equation}

Therefore, unlike in the spinless case, the transformations of the internal variables are not determined by a purely algebraic projection of $\Lambda_{AB}$. The WZ term deforms the spinless result by the first-order operator $1+\lambda m^Ae_A$.
For $s\neq 0$, the Noether equations imply that $\alpha$ must be Killing in the regular sector. For a given Killing vector, the remaining equations form an overdetermined first-order system for $\beta$ and $\gamma$.
 Since we are mainly interested in exhibiting the conformal transformations, we will not attempt a general study of this system.

   From the above discussion, it follows that the way to
   solve the NSE for $s=0$ is to find the (conformal) Killing vector
   fields of the spacetime metric, since this guarantees that the
   equations can be solved for $\beta$ and $\gamma$. We will proceed
   by going to the conformally flat coordinates, since the
   candidate fields are known for the flat metric. The conformal
   factor in the original coordinates can be obtained by direct
   computation using the push-forward vector field or using that if
   $G_{MN}=\Omega^2\eta_{MN}$, and \(\xi\) is a conformal Killing
   vector of the flat metric with $\mathcal
   L_\xi\eta_{MN}=2\sigma_\xi\eta_{MN}$, then
   \begin{equation}
     \mathcal L_\xi G_{MN}
     =
     2\left(\sigma_\xi+\xi(\log\Omega)\right)G_{MN}.
     \label{eq:LieConformal}
   \end{equation}
   In what follows we will give  the expressions only for the patch
   $x^2>0$, and we will use the shorthand $q\rho := \sqrt{q^2x^2}$.

\subsubsection*{Translations in the conformally flat coordinates}

We first consider translations in the conformally flat coordinates. Since
\(y^M=r(\rho)k^M\), with
\begin{equation}
	x^M=\rho k^M,
	\qquad
	\eta_{MN}k^Mk^N=1,
	\qquad
	r(\rho)=\frac{2}{q}\tanh\left(\frac{q\rho}{4}\right),
\end{equation}
we choose the normalization
\begin{equation}
	\delta_a y^M=\frac12 a^M,
	\label{eq:cf-translation-y}
\end{equation}
so that the flat limit gives \(\delta_a x^M\to a^M\). Here \(a^M\) is a
constant coordinate vector in the conformally flat chart. From
\begin{equation}
	y^2:=\eta_{MN}y^My^N=r^2,
\end{equation}
we get
\begin{equation}
	\delta_a r
	=
	\frac{y_M\delta_a y^M}{r}
	=
	\frac12 a\cdot k,
	\qquad
	a\cdot k:=\eta_{MN}a^Mk^N .
\end{equation}
Since \(\delta r=r'(\rho)\delta\rho\), this gives
\begin{equation}
	\delta_a\rho
	=
	\frac{a\cdot k}{2r'(\rho)} .
\end{equation}
Moreover,
\begin{equation}
	\delta_a k^M
	=
	\frac{1}{r}\left(\delta_a y^M-\delta_a r\,k^M\right)
	=
	\frac{1}{2r}\left(a^M-(a\cdot k)k^M\right).
\end{equation}
Therefore
\begin{align}
	\delta_a x^M
	&=
	\delta_a\rho\,k^M+\rho\,\delta_a k^M
	\nonumber\\
	&=
	\frac{a\cdot k}{2r'(\rho)}k^M
	+
	\frac{\rho}{2r(\rho)}
	\left(a^M-(a\cdot k)k^M\right).
	\label{eq:cf-translation-x-radial}
\end{align}
Equivalently, using \(k^M=x^M/\rho\),
\begin{equation}
	\delta_a x^M
	=
	\frac{\rho}{2r(\rho)}a^M
	+
	\left(
	\frac{1}{2\rho^2 r'(\rho)}
	-
	\frac{1}{2\rho r(\rho)}
	\right)
	(a\cdot x)x^M .
	\label{eq:cf-translation-x-coordinate}
\end{equation}
This is the conformal-vector field in the \(x^M\)-coordinates obtained by
pushing forward a translation in the conformally flat \(y^M\)-coordinates.
It is not, in general, the same as the teleparallel MB translation
\(a^Ae_A\).

The infinitesimal transformation generated by the constant parameter \(a^M\) defines the vector
field
\begin{equation}
X(a)
=
\left[
\frac{\rho}{2r(\rho)}a^M
+
\left(
\frac{1}{2\rho^2r'(\rho)}
-
\frac{1}{2\rho r(\rho)}
\right)
(a\cdot x)x^M
\right]\partial_M ,
\label{eq:cf-translation-x-field}
\end{equation}
where, for the conformally flat coordinates,
\begin{equation}
r(\rho)=\frac{2}{q}\tanh\left(\frac{q\rho}{4}\right).
\end{equation}
This vector field is the push-forward to the \(x^M\)-coordinates of a translation in the
conformally flat coordinates \(y^M=r(\rho)x^M/\rho\).  Since the metric can be written as
\begin{equation}
G_{MN}(x)\dd x^M\dd x^N
=
\Omega^2(\rho)\eta_{MN}\dd y^M\dd y^N,
\qquad
\Omega(\rho)=2\cosh^2\left(\frac{q\rho}{4}\right),
\end{equation}
and since translations are Killing vectors of \(\eta_{MN}\dd y^M\dd y^N\), the only variation comes
from the conformal factor.  Therefore \(X(a)\) is a conformal Killing vector of \(G_{MN}\),
\begin{equation}
\mathcal L_{X(a)}G_{MN}
=
2\Theta_a\,G_{MN},
\end{equation}
with
\begin{equation}
\Theta_a
=
\delta_a\log\Omega
=
\frac{q}{4\rho}
\sinh\left(\frac{q\rho}{2}\right)(a\cdot x).
\end{equation}
Thus \(X(a)\) is not a Killing vector for \(q\neq0\), because \(\Theta_a\) is then generically
nonzero, but it is a conformal Killing vector.  In the flat limit \(q\to0\), one has
\(\Theta_a\to0\), and the transformation reduces to an ordinary Killing translation.

\subsubsection*{Lorentz transformations}

For a Lorentz transformation in the conformally flat coordinates,
\begin{equation}
	\delta_\Lambda y^M=\Lambda^M{}_N y^N,
	\qquad
	\Lambda_{MN}=-\Lambda_{NM}.
	\label{eq:cf-lorentz-y}
\end{equation}
Since \(y^M=rk^M\), antisymmetry of \(\Lambda_{MN}\) implies
$
	\delta_\Lambda r
	=
	k_M\Lambda^M{}_N rk^N
	=0$,
and then
$
	\delta_\Lambda\rho=0$,
and
$	\delta_\Lambda k^M=\Lambda^M{}_N k^N$.
 It follows that the Lorentz vector field keeps the same form in the
\(x^M\)-coordinates:
\begin{equation}
	\delta_\Lambda x^M
	=
	\Lambda^M{}_N x^N,
	\label{eq:cf-lorentz-x}
\end{equation}
and the conformal factor is zero,
$
	\Theta_\Lambda=0$.

\subsubsection*{Dilatations}
\label{subsec:cf-dilatations}

For a dilatation in the conformally flat coordinates,
\begin{equation}
	\delta_D y^M=\epsilon_D y^M .
	\label{eq:cf-dilatation-y}
\end{equation}
Then
\begin{equation}
	\delta_D r=\epsilon_D r,
	\qquad
	\delta_D k^M=0.
\end{equation}
Since \(\delta r=r'(\rho)\delta\rho\), we get
\begin{equation}
	\delta_D\rho
	=
	\epsilon_D\frac{r(\rho)}{r'(\rho)} .
\end{equation}
Therefore
\begin{equation}
	\delta_D x^M
	=
	\epsilon_D\frac{r(\rho)}{r'(\rho)}k^M
	=
	\epsilon_D\frac{r(\rho)}{\rho r'(\rho)}x^M .
\end{equation}
Using
\begin{equation}
	r(\rho)=\frac{2}{q}\tanh\left(\frac{q\rho}{4}\right),
	\qquad
	r'(\rho)=\frac12\operatorname{sech}^2\left(\frac{q\rho}{4}\right),
\end{equation}
one obtains
\begin{equation}
	\frac{r(\rho)}{r'(\rho)}
	=
	\frac{2}{q}\sinh\left(\frac{q\rho}{2}\right).
\end{equation}
Hence
\begin{equation}
	\delta_D x^M
	=
	\epsilon_D
	\frac{2}{q\rho}
	\sinh\left(\frac{q\rho}{2}\right)x^M .
	\label{eq:cf-dilatation-x}
\end{equation}
The conformal factor is
\begin{equation}
	\Theta_D
	=
	\epsilon_D\cosh\left(\frac{q\rho}{2}\right).
	\label{eq:cf-dilatation-theta}
\end{equation}
In the flat limit \(q\to0\), this becomes
\begin{equation}
	\delta_D x^M\to \epsilon_D x^M,
	\qquad
	\Theta_D\to \epsilon_D .
\end{equation}

\subsubsection*{Special conformal transformations}

For a special conformal transformation in the conformally flat coordinates,
we use the normalization
\begin{equation}
	\delta_b y^M
	=
	-r^2 b^M+2y^M(b\cdot y),
	\qquad
	b\cdot y:=\eta_{MN}b^My^N,
	\label{eq:cf-special-y}
\end{equation}
where \(b^M\) is constant. Since \(y^M=rk^M\), this is
\begin{equation}
	\delta_b y^M
	=
	-r^2 b^M+2r^2Y k^M,
	\qquad
	Y:=b\cdot k .
\end{equation}
The radial variation is
\begin{equation}
	\delta_b r
	=
	k_M\delta_b y^M
	=
	r^2Y.
\end{equation}
Thus
\begin{equation}
	\delta_b\rho
	=
	\frac{r(\rho)^2}{r'(\rho)}Y .
\end{equation}
The angular variation is
\begin{align}
	\delta_b k^M
	&=
	\frac{1}{r}\left(\delta_b y^M-\delta_b r\,k^M\right)
	=
	-r b^M+rYk^M .
\end{align}
Therefore
\begin{align}
	\delta_b x^M
	&=
	\delta_b\rho\,k^M+\rho\,\delta_b k^M
=
	-\rho r(\rho)b^M
	+
	\left(
	\frac{r(\rho)^2}{r'(\rho)}+
	\rho r(\rho)
	\right)Yk^M .
\end{align}
Using
\begin{equation}
	Yk^M
	=
	\frac{x^M(b\cdot x)}{\rho^2},
\end{equation}
we can write the result as
\begin{equation}
	\delta_b x^M
	=
	A(\rho)b^M+B(\rho)x^M(b\cdot x),
	\label{eq:cf-special-x-general}
\end{equation}
where, using the explicit form of \(r(\rho)\),
\begin{equation}
	A(\rho)
	=
	-\frac{2\rho}{q}\tanh\left(\frac{q\rho}{4}\right),
\quad 
	B(\rho)
	=
	\frac{2}{q\rho}\tanh\left(\frac{q\rho}{4}\right)
	+
	\frac{8}{q^2\rho^2}
	\sinh^2\left(\frac{q\rho}{4}\right).
\end{equation}
Hence
\begin{equation}
	\delta_b x^M
	=
	-\frac{2\rho}{q}\tanh\left(\frac{q\rho}{4}\right)b^M
	+
	\left[
	\frac{2}{q\rho}\tanh\left(\frac{q\rho}{4}\right)
	+
	\frac{8}{q^2\rho^2}
	\sinh^2\left(\frac{q\rho}{4}\right)
	\right]
	x^M(b\cdot x).
	\label{eq:cf-special-x-explicit}
\end{equation}
The corresponding conformal factor is
\begin{equation}
	\Theta_b
	=
	\frac{2}{q\rho}
	\sinh\left(\frac{q\rho}{2}\right)(b\cdot x).
	\label{eq:cf-special-theta}
\end{equation}
In the flat limit \(q\to0\),
\begin{equation}
	A(\rho)\to -\frac12\rho^2,
	\qquad
	B(\rho)\to1,
\end{equation}
so that
\begin{equation}
	\delta_b x^M
	\to
	-\frac12\rho^2 b^M+x^M(b\cdot x),
	\qquad
	\Theta_b\to b\cdot x .
\end{equation}

\subsubsection*{From spacetime conformal vectors to internal transformations}

The previous subsections give the coordinate components of the conformal
vector fields in the \(x^M\)-coordinates. Thus the spacetime variation is
simply
\begin{equation}
	\delta x^M=\alpha^M(x).
	\label{eq:cf-delta-x-coordinate-final}
\end{equation}

To determine the induced transformations of the internal variables, one must compute
\begin{equation}
	K_{AB}=\Theta\eta_{AB}+\Lambda_{AB},
	\qquad
	\Lambda_{AB}=K_{[AB]} .
	\label{eq:cf-K-decomposition}
\end{equation}
For the spinless massless particle, the internal transformations are then
\begin{equation}
	\delta u
	=
	-m^A\Lambda^B{}_A n_B,
	\qquad
	\delta\varphi
	=
	\ell^A\Lambda^B{}_A n_B-\Theta .
	\label{eq:cf-spinless-internal-transformations}
\end{equation}

In light-cone components, the same formulae may be written as
\begin{equation}
	\delta u
	=
	-\Lambda_{+2}-u\Lambda_{+-}-\frac12u^2\Lambda_{2-},
	\label{eq:cf-delta-u-lightcone}
\end{equation}
and
\begin{equation}
	\delta\varphi
	=
	\Lambda_{+-}+ u \Lambda_{2-}-\Theta .
	\label{eq:cf-delta-varphi-lightcone}
\end{equation}

For Lorentz transformations, \(\Theta=0\) and \(\Lambda_{AB}\) is just the
constant Lorentz parameter written in the local frame. Hence
\begin{equation}
	\delta_\Lambda u
	=
	-m^A\Lambda^B{}_A n_B,
	\qquad
	\delta_\Lambda\varphi
	=
	\ell^A\Lambda^B{}_A n_B .
\end{equation}

 For the  dilatation in \eqref{eq:cf-dilatation-x}, 
one can show, using (\ref{eq:KAB4}) and (\ref{eq:app-frame-inverse-components2}), that
\begin{equation}
	K_{AB} =\epsilon_D \cosh\frac{q\rho}{2} \eta_{AB} + \epsilon_D \frac{\sinh(q\rho/2)}{\rho}\epsilon_{ABC} x^C
	\label{eq:KABD}
\end{equation}	
with a nonzero antisymmetric part that contributes to the transformation of $u$ and $\varphi$.	

Similarly,  $K_{AB}$ can also be computed for translations and special conformal transformations, but we do not provide the details here.

	\subsection{Conformal MB algebra of the spinless massless particle}
	\label{sec:algebra}
	
	In this subsection we return exclusively to the spinless massless
	particle, $s=0$. As shown in the previous subsection, all conformal
	Killing vectors of the MB metric lift to Noether transformations of the
	spinless action. The resulting symmetry algebra is therefore the
	three-dimensional conformal algebra $\mathfrak{so}(3,2)$.
	
	There are, however, two different natural notions of translation in the
	present geometry. The first consists of the ordinary translations in
	the conformally flat coordinates introduced in~\ref{sec:conformal-symmetry}. The second consists of
	the translations belonging to the MB isometry algebra. These two sets
	of generators coincide in the flat limit, but differ when $q\neq0$.
	The purpose of this subsection is to explain their relation and to
	rewrite the conformal algebra in a basis adapted to the teleparallel MB
	isometries.
	
	\subsubsection*{Conformal-coordinate translations.}

	The conformal Killing vectors that we have obtained using the conformally flat coordinates are the standard generators of
	the conformal algebra in three-dimensional Minkowski space. We denote
	them by
	$
		\left\{
		J_A,\Pi_A,D,K_A
		\right\}$,
	where $J_A$ generate Lorentz transformations, $\Pi_A$ ordinary
	translations in the $y^M$ coordinates, $D$ dilatations and $K_A$
	special conformal transformations.

	For constant Lorentz vectors $a^A$, $b^A$ and $\lambda^A$, we use the
	notation
	\begin{equation}
		\Pi(a)=a^A\Pi_A,
		\qquad
		J(\lambda)=\lambda^A J_A,
		\qquad
		K(b)=b^A K_A,
		\label{eq:smeared-standard-conformal-generators}
	\end{equation}
	together with $(a\times b)^A=\epsilon^A{}_{BC}a^Bb^C$.
	
	In terms of commutators of vector fields, the standard conformal algebra is
	\begin{equation}
		\begin{aligned}
			[J(\lambda_1),J(\lambda_2)]
			&=
			-J(\lambda_1\times\lambda_2),
			\\
			[J(\lambda),\Pi(a)]
			&=
			-\Pi(\lambda\times a),
			\\
			[J(\lambda),K(b)]
			&=
			-K(\lambda\times b),
			\\
			[J(\lambda),D]
			&=
			0,
		\end{aligned}
		\label{eq:standard-conformal-Lorentz-sector}
	\end{equation}
	and
	\begin{equation}
		\begin{aligned}
			[D,\Pi(a)]
			&=
			-\Pi(a),
			\\
			[D,K(b)]
			&=
			K(b),
			\\
			[\Pi(a_1),\Pi(a_2)]
			&=
			0,
			\\
			[K(b_1),K(b_2)]
			&=
			0,
			\\
			[\Pi(a),K(b)]
			&=
			(a\cdot b)D+J(a\times b).
		\end{aligned}
		\label{eq:standard-conformal-non-Lorentz-sector}
	\end{equation}
	
	The generators $\Pi_A$ are translations only with respect to the
	conformally flat coordinates $y^M$. When pushed forward to the original
	coordinates $x^M$, they become the nonlinear conformal vector fields (\ref{eq:cf-translation-x-field}) . They are
	not, for $q\neq0$, isometries of the MB metric. Instead, they satisfy
	\begin{equation}
		\mathcal{L}_{\Pi(a)}G
		=
		2\Theta_{\Pi(a)}G,
		\label{eq:conformal-translations-Weyl-factor}
	\end{equation}
	with a generally nonzero conformal factor
	$\Theta_{\Pi(a)}$.
	
	\subsubsection{Teleparallel MB translations}
	
	In the teleparallel geometry there exists a preferred  coframe which, in the coordinate patch defined by the coset representative $g_0(x)=e^{x^AP_A}$ is
	$g_0^{-1}dg_0=\theta^A P_A$.
	The vector fields dual to the coframe,
	\begin{equation}
		\theta^A(P_B)=\delta^A{}_B,
		\label{eq:thetaP}
	\end{equation}
	will be referred to as the \emph{teleparallel MB translation generators}.
	Notice that these generators are left-invariant vector fields on the
	three-dimensional translation subgroup.
	
	The Maurer--Cartan equation (\ref{eq:MC}) immediately implies
	$-\theta^A([P_B,P_C])
		=
		d\theta^A(P_B,P_C)
		\nonumber\\
		=
		-q\epsilon^A{}_{BC}$,
	so that
	\begin{equation}
		[P_A,P_B]
		=
		q\epsilon_{AB}{}^{C}P_C.
		\label{eq:MBtranslationalg}
	\end{equation}
	
 	These generators are Killing vectors. Indeed, using the Maurer-Cartan equation (\ref{eq:MC}) and (\ref{eq:thetaP})  one immediately gets $\mathcal L_{P_B} \theta^A = -q \epsilon^A{}_{BC}\theta^C$. This is a Lorentz rotation of the coframe, and hence  $\mathcal L_{P_B}G=0$.
		Unlike the ordinary coordinate translations
	$\Pi_A=\partial/\partial y^A$
	introduced in the conformally-flat coordinates,
	the generators $P_A$ are adapted to the MB geometry itself.
		We now determine their relation with the standard conformal basis
	$(\Pi_A,J_A,K_A)$.
	
	Lorentz covariance implies that $P_A$ must be a constant linear
	combination of the three Lorentz-vector generators,
	\begin{equation}
		P_A
		=
		\Pi_A+\alpha J_A+\beta K_A,
		\label{eq:PBansatz}
	\end{equation}
	where the normalization of $\Pi_A$ has been chosen so that
	$P_A\rightarrow\Pi_A$ in the flat limit $q\rightarrow0$.
	
	The coefficients $\alpha$ and $\beta$ are fixed by requiring that
	(\ref{eq:PBansatz}) satisfy the MB algebra
	(\ref{eq:MBtranslationalg}).
	Using the conformal commutation relations,
	one finds
	\begin{equation}
		\alpha=-\frac q2,
		\qquad
		\beta=-\frac{q^2}{8},
	\end{equation}
	and therefore
	\begin{equation}
			P_A
			=
			\Pi_A
			-\frac q2J_A
			-\frac{q^2}{8}K_A.
		\label{eq:MBleftbasis}
	\end{equation}
	
	A direct computation verifies that
	$
		[P_A,P_B]
		=
		q\epsilon_{AB}{}^CP_C
	$,
	as required.
	
	The appearance of the Lorentz and special-conformal generators reflects
	the fact that the teleparallel translations are adapted to the
	non-trivial geometry rather than to the conformally-flat coordinates.
	In the limit $q\rightarrow0$,
	(\ref{eq:MBleftbasis}) reduces smoothly to
	$P_A\longrightarrow\Pi_A$,
	recovering ordinary Minkowski translations.
	
	It should be stressed that the above generators are the
	left-invariant vector fields dual to the Maurer--Cartan coframe.
	Equally natural are the right-invariant Killing vectors, which generate
	the left action of the translation subgroup on itself and satisfy the
	opposite Lie algebra,
	\begin{equation}
		[\widehat P_A,\widehat P_B]
		=
		-q\epsilon_{AB}{}^C\widehat P_C.
	\end{equation}
	Both triples are Killing and coincide at the group identity, but differ
	away from it.  The present discussion employs the left-invariant
	realization because it is naturally associated with the teleparallel
	coframe.
	
	\subsubsection{Conformal algebra in the left-invariant teleparallel basis}
	
	If we replace the conformal-coordinate translations \(\Pi_A\) by the left-invariant
	teleparallel MB translations
	\begin{equation}
		P_A=\Pi_A-\frac q2 J_A-\frac{q^2}{8}K_A,
		\label{eq:left-MB-basis-change}
	\end{equation}
	or equivalently
	\begin{equation}
		\Pi_A=P_A+\frac q2 J_A+\frac{q^2}{8}K_A,
		\label{eq:left-MB-inverse-basis-change}
	\end{equation}
the conformal algebra in the basis
	\(\{J_A,P_A,D,K_A\}\) is
	\begin{align}
		[J(\lambda_1),J(\lambda_2)]
		&=-J(\lambda_1\times\lambda_2),
		\\
		[J(\lambda),P(a)]
		&=-P(\lambda\times a),
		\\
		[J(\lambda),K(b)]
		&=-K(\lambda\times b),
		\\
		[J(\lambda),D]&=0,
		\\
		[P(a_1),P(a_2)]
		&=q\,P(a_1\times a_2),
		\\
		[D,P(a)]
		&=-P(a)-\frac q2J(a)-\frac{q^2}{4}K(a),
		\\
		[D,K(b)]
		&=K(b),
		\\
		[P(a),K(b)]
		&=(a\cdot b)D+J(a\times b)
		+\frac q2K(a\times b),
		\\
		[K(b_1),K(b_2)]
		&=0,
	\end{align}
	where $P(a)=a^AP_A$.
	The explicit \(q\)-dependence does not represent any deformation and is only a consequence of the choice of basis.
	Indeed, the invertible change of basis
	\eqref{eq:left-MB-inverse-basis-change} brings the algebra back to the standard
	conformal form, so the Lie algebra remains isomorphic to
	\(\mathfrak{so}(3,2)\).
	
	The advantage of the teleparallel basis is that the six generators
	$
	\left\{
	J_A,P_A
	\right\}
	$
	form a manifest MB isometry subalgebra. The remaining generators
	$D$ and $K_A$ are proper conformal generators. This distinction is
	particularly useful when comparing with the spinning theory. The
	Wess--Zumino term obstructs the proper conformal transformations, while
	the MB isometries generated by $J_A$ and $P_A$ remain genuine Noether
	symmetries. In the critical spinning sector, the non-Killing spacetime contribution is gauge removable, leaving a Killing vector. 
	Since the full Killing algebra of the MB metric is the six-dimensional MB isometry algebra spanned by $J_A$ and $P_A$, the resulting transformation is gauge-equivalent to an MB isometry.

\section{Conclusions and outlook}
\label{sec:concl}

We have studied the geometry and dynamical realizations of Mielke-Baekler (MB) spacetimes in $2+1$ dimensions, which are controlled by two parameters $p$ and $q$.

We have constructed  particle actions   from coadjoint orbits via the method of nonlinear
realisations. The same procedure determines the background Cartan
data, yielding homogeneous Riemann--Cartan spacetimes with invariant
 torsion and curvature.

Although we have described some general results for the MB spacetime
for general values of the parameters $p$ and $q$, the particle actions
have been constructed for the teleparallel $p=0$ case, since it allows
a dual description in terms of the Weitzenböck and Levi-Civita
connections.    In this last case we have also shown that
  the metric can be re-expressed as a manifestly conformally flat
  metric by performing a suitable transformation on the space-time
  coordinates.

Both massive and massless particles have been considered, and for each
of them we have studied the effect of spin, introduced via a WZ
term. This term produces the appearance of different dynamical sectors,
which have different physical degrees of freedom and dynamics.

In the regular sectors, \(m+qs\neq0\) for the massive
particle and \(\mu +qs\neq0\) for the massless particle, the internal
Goldstone variables are removed by second-class constraints and do not give rise
to additional propagating degrees of freedom.  Moreover, once the regular-sector
constraints are imposed, the spacetime equations reduce to the ordinary
geodesic equations: timelike geodesics in the massive case and null geodesics in
the massless case.  Thus, at the level of unparametrised spacetime trajectories,
the regular spinning particles behave as their spinless counterparts.

This does not mean, however, that the Wess--Zumino terms are
dynamically irrelevant. Their coupling constants, \(s\) in the massive
case and \(s\) in the massless case, survive the elimination of the
internal variables through the reduced symplectic structure. In
particular, the regular-sector Dirac brackets of the spacetime
coordinates acquire spin-dependent contributions. Thus, the
Wess--Zumino terms do not introduce extra local propagating degrees of
freedom in the regular sectors, nor do they produce a force term in
the spacetime equations of motion, but they do deform the reduced
phase-space geometry. This distinction is important for the canonical
description, for the realisation of the conserved charges and,
ultimately, for quantisation.

The critical sectors have a qualitatively different interpretation.  At the
critical loci, \(m+qs=0\) for the massive particle and \(\mu+qs=0\) for the
massless particle, the presymplectic form degenerates further.  Equivalently,
the matrix of Poisson brackets of the internal constraints drops rank, and two
constraints which are second class in the regular sector become first class.
This produces two additional gauge symmetries, acting on the internal Goldstone
variables together with compensating transformations of the spacetime
coordinates.  As a result, the dimension of the physical phase space is reduced:
in both the massive and massless cases the regular four-dimensional reduced
phase space collapses to a two-dimensional one.  This agrees with the
corresponding coadjoint-orbit picture, where the generic four-dimensional orbit
drops to a two-dimensional orbit at the exceptional value of the orbit
parameters.

In these critical sectors the Wess--Zumino coupling is no longer merely a
deformation of the reduced symplectic structure.  It controls the very
degeneracy of the presymplectic form and hence the emergence of the extra gauge
invariance.  Correspondingly, the momentum--velocity relation becomes
degenerate: the momentum fixes only part of the worldline velocity, while the
remaining components are gauge.  In the massive case this is the sector in which
the Levi--Civita description acquires a genuine Papapetrou-type spin--curvature
force.  In the massless case the same mechanism appears through the enlarged
first-class constraint set generated by the critical relation
\(\mu=-qs\).  Thus the critical branches are not obtained by a smooth
elimination of auxiliary spin variables from the regular theory; rather, they
represent distinct constrained systems with fewer physical degrees of freedom
and enhanced gauge symmetry.

In the massless case we have also investigated the presence of
generalized conformal symmetries. We have first analyzed the general
structure of the equations for the symmetry (Noether) generators and
shown that these equations admit conformal solutions only when $s$,
the parameter associated to the Wess--Zumino term, vanishes. To be
more specific, for $s\neq 0$, in the regular sector, \(\mu +qs\neq
0\), the NSE force the transformations to be Killing, and in the
critical sector, \(\mu +qs=0\), the conformal Killing contributions
can be completely removed by gauge transformations.

Starting from dilation and special conformal transformations in the
conformally flat metric space and  making use of the relevant
transformation,  we have derived the generalized conformal
transformations in the original space-time coordinates. Finally,
always in the spinless case, we have analyzed the relation between the
standard conformal algebra and the MB  algebra in the teleparallel
basis. In particular we have derived the relation between the
translation generators in the two bases: it turns out that  the
teleparallel MB translations are not the ordinary translations, but
they contain both a Lorentz and a special conformal generator
contribution.

Possible extensions of the work presented in this paper include the
study of strings in torsional MB backgrounds, the carrollian limits \cite{Concha:2025vhd}
and supersymmetric versions of the MB spacetimes and particles
\cite{Giacomini:2006dr,Cvetkovic:2007sr,Barriga:2026awj}.

\acknowledgments We thank Nelson Merino for useful discussions and for
pointing out several references. JG acknowledges the hospitality and
support of the Galileo Galilei Institute, where part of this work was
done.

CB was supported in part by DECODER project (PID2024-158394OB-C22),
funded by the Spanish Ministry of Science, Innovation and
Universities, the Spanish National Research Agency and the European
Regional Development Fund (MICIU/AEI/10.13039/501100011033/FEDER UE).
JG acknowledges financial support from the Spanish
MCIN/AEI/10.13039/501100011033 grant PID2022-126224NB-C21.

\appendix

\begin{appendices}

\section{Re-interpretation of the MB-algebra in terms of deformations of Lie algebras}
\label{sec:deform-lie-algebr}

In this Appendix we observe how the Nomizu map and the Lie algebra
\eqref{eq:MB-algebra} appear from the point of view of deformation of
Lie algebras.

We depart from the three-dimensional Poincaré Lie algebra $\g = \h
\ltimes \fm$, where $\h \cong \so(2,1)$ and $\fm \cong \RR^3$ is
abelian.  We will choose a basis $J_A$ for $\h$ and $P_A$ for $\fm$,
with $A\in\{0,1,2\}$, where the Lie brackets are given by
\begin{equation}
  \label{eq:Poincare}
  \begin{split}
    [J_A, J_B] &= \epsilon_{ABC} J^C\\
    [J_A, P_B] &= \epsilon_{ABC} P^C\\
    [P_A, P_B] & = 0,
  \end{split}
\end{equation}
with $\epsilon_{ABC}$ the totally antisymmetric Levi-Civita symbol
normalised to $\epsilon_{012}=+1$ and where we raise and lower indices
with the lorentzian inner product $\eta_{AB} =
\operatorname{diag}(-1,1,1)$.

The dual description of this Lie algebra is as the
Chevalley--Eilenberg \cite{ChevalleyEilenberg} differential $d : \g^*
\to \ext{2}\g^*$, which is the dual (i.e., transpose) of the Lie
bracket $\ext{2}\g \to \g$.  Let us introduce the canonical dual basis
$\lambda^A, \pi^A$ for $\g^*$, where the dual pairing is given by
\begin{equation}
  \label{eq:duality}
  \left<\lambda^A, J_B\right> = \delta^A_B,\qquad \left<\pi^A,
    P_B\right> = \delta^A_B, \qquad \left<\lambda^A, P_B\right> =
  \left<\pi^A, J_B\right> = 0.
\end{equation}
If $\alpha \in \g^*$, then $d\alpha \in\ext{2}\g^*$ is defined by
\begin{equation}
  \label{eq:cobracket}
  d\alpha(X,Y) = - \left<\alpha, [X,Y]\right>,
\end{equation}
from where we see that
\begin{equation}
  \label{eq:dual-lie-algebra}
  d \lambda_C = - \tfrac12 \epsilon_{ABC} \lambda^A \wedge \lambda^B
  \qquad\text{and}\qquad d\pi_C = - \epsilon_{ABC} \lambda^A \wedge  \pi^B.
\end{equation}
Since $\h$ is simple, the Hochschild--Serre factorisation theorem
\cite{MR0054581} says that equivalence classes of infinitesimal
deformations of $\g$ are given by $H^2(\fm,\g)^\h$; that is, the
$\h$-invariant cohomology of the abelian Lie algebra $\fm$ with values
in the representation $\g$.  Since $\h$ is simple, it acts reducibly
on the cochains and hence the $\h$-invariant cohomology can be
calculated from the $\h$-invariant cochains.

Since $\dim \fm = 3$ we have $p$-cochains for $p=0,1,2,3$.  However
there are no $\h$-invariant $0$- and $3$-cochains.  The $\h$-invariant
$0$-cochains would be $\h$-invariant elements in $\g$, but there are
none: both $J_A$ and $P_A$ transform as vectors under $\h$ and there
are no nonzero invariant vectors.  The $\h$-invariant $3$-cochains are
$\h$-equivariant maps $\ext{3}\fm \to \g$, but $\ext{3}\fm$ is a
one-dimensional trivial $\h$-representation and there are no
$\h$-invariants in $\g$.  Therefore we have a two-step complex
\begin{equation}
  \label{eq:CE-complex-deformations}
  \begin{tikzcd}
    C^1(\fm;\g)^\h \arrow[r,"d"] & C^2(\fm;\g)^\h.
  \end{tikzcd}
\end{equation}
The space of $\h$-invariant $1$-cochains is two-dimensional,
consisting of linear maps $\lambda : \fm \to \g$ given by
\begin{equation}
  \label{eq:1-cochains}
  \lambda(P_A) = s P_A + t J_A
\end{equation}
whereas the space of $\h$-invariant $2$-cochains, which is also
two-dimensional, consists of linear maps $\mu : \ext{2}\fm \to \g$
given by
\begin{equation}
  \label{eq:2-cochains}
  \mu(P_A,P_B) = \epsilon_{ABC} (p J^C + q P^C).
\end{equation}
The Chevalley--Eilenberg differential on the $1$-cochains $\lambda$ is
such that
\begin{equation}
  \label{eq:2-coboundaries}
  \begin{split}
    (d \lambda)(P_A,P_B) &= [P_A, \lambda(P_B)] - [P_B, \lambda(P_A)] - \lambda([P_A,P_B])\\
    &= [P_A, s P_B + t J_B] - [P_B, s P_A + t J_A]\\
    &= -t [J_B, P_A] + t [J_A, P_B]\\
    &= 2 t \epsilon_{ABC} P^C.
  \end{split}
\end{equation}

We notice that (perhaps coincidentally due to the paucity of invariant
maps in this case), the expression \eqref{eq:2-cochains} of the
$\h$-invariant $2$-cocycles precisely agree (up to an inconsequential
sign) with the Lie bracket $[P_A,P_B]$ in
equation~\eqref{eq:MB-algebra} and the expression
\eqref{eq:2-coboundaries} for the $\h$-invariant $2$-coboundaries
agree (up to an inconsequential factor of $2$) with the Nomizu map
\eqref{eq:Nomizu}.

The infinitesimal deformations are then given by
\begin{equation}
  \label{eq:inf-defs}
  H^2(\g;\g) \cong H^2(\fm;\g)^\h  = \frac{\displaystyle
    C^2(\fm;\g)^\h}{d C^1(\fm;\g)^\h} = \frac{\displaystyle
    \{\mu(P_A,P_B) = \epsilon_{ABC} (p J^C + q P^C)\}}{\displaystyle
    \{d\lambda(P_A,P_B) = 2t \epsilon_{ABC} P^C\}},
\end{equation}
so that we can set $q$ to any fixed value we desire, for example
$q=0$.  There are no obstructions to integrating the infinitesimal
deformations.  Indeed, one can check that the Lie brackets
corresponding to any $\h$-invariant $2$-cocycle automatically satisfy
the Jacobi identity.

Finally, we may rescale the generators $P_A$ to further set $p=0,1,-1$
corresponding, respectively, to the Poincaré algebra, $\so(2,2)$ and
$\so(3,1)$.

\section{Teleparallel geometry in brief}
\label{sec:teleparallelreview}

Teleparallel geometry is the underpinning mathematical structure for
teleparallel gravity: a formulation of gravity in which the
fundamental object is a global frame.  See
\cite{Aldrovandi:2013wha,Pereira:2019} for introductions.

The existence of a global frame requires the spacetime to have trivial
tangent bundle; that is, that it should be parallelisable.  There are
topological obstructions to the triviality of the tangent bundle of a
manifold, so not every manifold is parallelisable.  However in three
dimensions, orientability is enough to guarantee it.

Every global frame $e_A$ defines a metric $G$ by declaring the vector
fields in the frame to be (pseudo)orthonormal:
$G(e_A,e_B) = \eta_{AB}$.  Equivalently, if we let $\theta^A$ denote
the canonical dual coframe, the metric is given by $G = \eta_{AB}
\theta^A \theta^B$.

Similarly, every global frame defines a flat affine connection
$\nabla$ by declaring the vector fields in the frame to be parallel:
$\nabla e_A = 0$.  Since $\eta_{AB}$ are constant, it follows that
$\nabla$ is metric compatible: $\nabla G = 0$.  Since $G$ need not be
flat, whereas $\nabla$ is, it must have torsion generically.  The
connection $\nabla$ is called a Weitzenböck connection.

Let us choose local coordinates $x^M$, relative to which $e_A = e_A^M
\partial_M$ and $\theta^A = \theta^A_M \dd x^M$.  The metric has
components $G_{MN} = \eta_{AB} \theta^A_M \theta^B_N$.  Defining the
connection coefficients $\Gamma_{MN}^P$ by
\begin{equation}
  \nabla_M \partial_N = \Gamma_{MN}^P \partial_P,
\end{equation}
and using that $\nabla_M e_A = 0$, we see that
\begin{equation}
  \label{eq:Weitz-conn-coeff}
  \Gamma_{MN}^P = - \theta^A_N \partial_M e_A^P = e_A^P \partial_M
  \theta^A_N.
\end{equation}
The torsion tensor is given as usual by the skew-symmetrisation of the
connection coefficients:
\begin{equation}
  \label{eq:Torsion-Weitzenboeck}
  T_{MN}^P = \Gamma_{MN}^P - \Gamma_{NM}^P = (\partial_M \theta^A_N - \partial_N \theta^A_M) e_A^P.
\end{equation}
If the torsion were zero, then $\dd\theta^A =0$ and hence, by the
Poincaré Lemma, locally there exist functions $y^A$ such that
$\theta^A = \dd y^A$.  These $y^A$ define local coordinates relative to
which the metric has constant coefficients and hence the metric is
flat.  Therefore the torsion of the Weitzenböck connection is the
obstruction to flatness of the metric.

The difference between any two connections is a tensor.  The
difference between a metric connection with torsion and the
Levi-Civita connection is the contorsion tensor $K$ and it is
determined by the torsion.  This follows from the standard fact that a
metric connection is uniquely specified by its torsion.  For a
metric connection $\nabla$, the contorsion tensor is defined for all
vector fields $X,Y$ by
\begin{equation}
  K(X,Y) = \nabla_X Y - \nablag_X Y,
\end{equation}
with $\nablag$ the Levi-Civita connection.  In terms of the connection
coefficients,
\begin{equation}
  K_{MN}^P = \Gamma_{MN}^P -\widetilde{\Gamma}_{MN}^P,
\end{equation}
with $\widetilde{\Gamma}_{MN}^P$ the Christoffel symbols of the
Levi-Civita connection.  Because $\nabla$ is metric-compatible, it
follows that
\begin{equation}
  K_{PM}^Q G_{QN} + K_{PN}^Q G_{QM}  = 0.
\end{equation}
Let $K_{MNP}:= K_{MN}^Q G_{QP}$.  Then metric-compatibility says
that $K_{MNP} = - K_{MPN}$.  Since $\nablag$ is torsion-free, we also
have that
\begin{equation}
  T_{MN}^P = K_{MN}^P - K_{NM}^P.
\end{equation}
These two conditions are enough to solve for $K_{MN}^P$ and we find
that
\begin{equation}
  K_{MNP} = \tfrac12 \left( T_{MNP} + T_{PNM} + T_{PMN}\right),
\end{equation}
where $T_{MNP} := T_{MN}^Q G_{QP}$.

Affinely-parametrised geodesics of the Weitzenböck connection are
generally not geodesics of the Levi-Civita connection, but contain
a torsional force term.  Indeed, an affinely-parametrised geodesic of
$\nabla$ obeys
\begin{equation}
  \ddot x^P + \Gamma_{MN}^P \dot x^M \dot x^N = 0 \implies \ddot x^P
  + \widetilde{\Gamma}_{MN}^P \dot x^M \dot x^N + K_{MN}^P \dot x^M \dot x^N,
\end{equation}
so that
\begin{equation}
  \ddot x^P  + \widetilde{\Gamma}_{MN}^P \dot x^M \dot x^N = -\tfrac12 (
  T_{MNQ} + T_{QNM} + T_{QMN} ) G^{QP} \dot x^M \dot x^N.
\end{equation}

\section{Gauge fixing in the regular and critical sectors for the massive particle}
\label{sec:gf}

In the regular sector the only gauge freedom is the
worldline reparametrisation generated by \(C\), and we can remove it by imposing the time gauge
\begin{equation}
	\label{eq:appC01}
	x^0=\tau,
\end{equation}
which turns \(C\approx0\) into  a second-class constraint.
Since \(\dot x^0=1\), the gauge-fixed lagrangian takes the form
\begin{equation}
	\label{eq:appC03}
	L_{\rm gf}=p_i\dot x^i+p_0
	=
	p_i\dot x^i-H_{\rm reg},
\end{equation}
and hence
\begin{equation}
	\label{eq:appC04}
	H_{\rm reg}=-p_0 .
\end{equation}
Here \(p_0\) is determined by the mass-shell constraint
\begin{equation}
	\label{eq:appC05}
	G^{00}p_0^2 + 2 G^{0i} p_0p_i + G^{ij}p_ip_j + m^2=0,
\end{equation}
with the branch solution  chosen so
that \(H_{\rm reg}=-p_0\) has the appropriate sign for the physical energy.

After imposing the second-class constraints \(\Theta_I\), the Dirac
bracket among the spacetime coordinates is
\begin{equation}
	\label{eq:appC07}
	\{x^M,x^N\}_D
	=
	-\frac{s}{m(m+qs)}
	\epsilon_{ABC}u^C e^M_A e^N_B .
\end{equation}
Using the constraint \(p_A=-m u_A\), this may also be written as
\begin{equation}
	\label{eq:appC08}
	\{x^M,x^N\}_D
	=
	\frac{s}{m^2(m+qs)}
	\epsilon_{ABC}p^C e^M_A e^N_B :=B^{MN}.
\end{equation}
The further imposition of the time gauge \(x^0=\tau\) modifies the bracket by the
standard Dirac projection with respect to the pair \((C,x^0-\tau)\). For any two
functions \(F,G\) on the regular reduced phase space,
\begin{equation}
	\label{eq:appC10}
	\{F,G\}_{\rm reg}
	=
	\{F,G\}_D
	-
	\frac{\{F,C\}_D\{x^0,G\}_D}{\{x^0,C\}_D}
	+
	\frac{\{F,x^0\}_D\{C,G\}_D}{\{x^0,C\}_D}.
\end{equation}
The gauge is admissible whenever $\{x^0,C\}_D\neq0$.
Writing $
	V^M:=\{x^M,C\}_D$,
the spatial coordinate brackets become
\begin{equation}
	\label{eq:appC13}
	\{x^i,x^j\}_{\rm reg}
	=
	B^{ij}
	+
	\frac{
		V^jB^{0i}-V^iB^{0j}
	}{V^0}.
\end{equation}
Equivalently, since on the reduced surface
$
	V^M=2g^{MN}p_N$,
one may write
\begin{equation}
	\label{eq:appC15}
	\{x^i,x^j\}_{\rm reg}
	=
	B^{ij}
	+
	\frac{
		G^{jN}p_N B^{0i}
		-
		G^{iN}p_N B^{0j}
	}{
		G^{0N}p_N
	} .
\end{equation}
In particular, the bracket between the two spatial coordinates is
\begin{equation}
	\label{eq:appC16}
	\{x^1,x^2\}_{\rm reg}
	=
	B^{12}
	+
	\frac{
		G^{2N}p_N B^{01}
		-
		G^{1N}p_N B^{02}
	}{
		G^{0N}p_N
	}.
\end{equation}

The resulting regular-sector reduced phase space is four-dimensional, coordinatised for
example by \((x^i,p_i)\), \(i=1,2\).

In the critical sector
the first-class constraints are \(C,\Gamma_v,\Gamma_\varphi\), while
\(\chi_v,\chi_\varphi\) may be kept as a second-class pair.  We fix
the three first-class freedoms locally by imposing the gauge
conditions
\begin{equation}
	\label{eq:appC18}
	x^0=\tau,\qquad v=\hat{v} ,\qquad \varphi=0 ,
	\qquad \hat{v}\neq0 .
\end{equation}

The constraints \(\Phi_A\approx0\) imply
$p_A=-m \hat{u}_{A}$ with $\hat{u}_{A}=(\cosh \hat{v},\sinh \hat{v},0)$
and therefore, in spacetime components,
\begin{equation}
	\label{eq:appC22}
	p_M=-m \hat{u}_{A}\theta^A{}_M
	=
	-m\left(\cosh \hat{v}\,\theta^0{}_M+\sinh \hat{v}\,\theta^1{}_M\right).
\end{equation}

The gauge-fixed lagrangian is obtained from the first-order form $L=p_M\dot x^M+s(\cosh v-1)\dot\varphi$.
Since \(x^0=\tau\) and \(\dot\varphi=0\), this becomes
\begin{equation}
	\label{eq:appC24}
	L_{\rm gf}
	=
	p_i\dot x^i+p_0,
	\qquad i=1,2 .
\end{equation}
Writing it in the standard form
\begin{equation}
	\label{eq:appC25}
	L_{\rm gf}
	=
	a_i(\tau,x)\dot x^i-H_{\rm red}(\tau,x),
\end{equation}
we identify
\begin{equation}
	\label{eq:appC26}
	a_i=p_i
	=
	-m\left(\cosh \hat{v}\,\theta^0{}_i+\sinh \hat{v}\,\theta^1{}_i\right),
\end{equation}
and
\begin{equation}
	\label{eq:appC27}
	H_{\rm red}=-p_0
	=
	m\left(\cosh \hat{v}\,\theta^0{}_0+\sinh \hat{v}\,\theta^1{}_0\right)
	\bigg|_{x^0=\tau}.
\end{equation}

Let
\begin{equation}
	\label{eq:appC29}
	R:=\left. x^2\right|_{x^0=\tau}=-\tau^2+(x^1)^2+(x^2)^2,
	\qquad
	B:=\frac{1-S}{R},
\end{equation}
so that \(S=S(R)\) and \(C=C(R)\). Then, using the explicit teleparallel coframe in (\ref{eq:theta-and-omega}),
\begin{equation}
	\label{eq:appC30}
	\theta^0{}_0=S-B\tau^2,
	\qquad
	\theta^1{}_0=-B\tau x^1+qC x^2 ,
\end{equation}
and thus
\begin{equation}	
	\label{eq:appC31}
	H_{\rm red}
	=
	m\left[
	\cosh \hat{v}\left(S-B\tau^2\right)
	+
	\sinh \hat{v}\left(-B\tau x^1+qC x^2\right)
	\right] .
\end{equation}
Equivalently, using \(m=-qs\),
\begin{equation}
	\label{eq:appC32}
	H_{\rm red}
	=
	-qs\left[
	\cosh \hat{v}\left(S-B\tau^2\right)
	+
	\sinh \hat{v}\left(-B\tau x^1+qC x^2\right)
	\right] .
\end{equation}

The reduced symplectic form is
\begin{equation}
	\label{eq:appC33}
	\Omega_{\rm red}
	=
	\frac12\Omega_{ij}\,\dd x^i\wedge \dd x^j,
	\qquad
	\Omega_{ij}
	=
	\partial_i a_j-\partial_j a_i,
\end{equation}
and using the Maurer--Cartan equation (\ref{eq:MC})
one obtains
\begin{equation}
	\label{eq:appC35}
	\Omega_{\rm red}
	=
	m q\,D\,\dd x^1\wedge \dd x^2 ,
\end{equation}
where
\begin{equation}
	\label{eq:appC36}
	D
	:=
	\hat{u}_{A}\epsilon^A{}_{BC}\theta^B{}_1\theta^C{}_2 .
\end{equation}
In the present gauge this is
\begin{equation}
	\label{eq:appC37}
	D
	=
	-\cosh \hat{v}\left(S+\tau^2 A\right)
	-
	\sinh \hat{v}\left(\tau x^1 A+qC x^2\right),
\end{equation}
with $A:=SB+q^2C^2$.
The gauge is  locally admissible whenever $D\neq0$.

The Dirac bracket on the reduced phase space is the inverse of the reduced symplectic
form. Thus, for two functions \(F(x^1,x^2)\) and \(G(x^1,x^2)\),
\begin{equation}
	\label{eq:appC40}
	\{F,G\}_{\rm red}
	=
	-\frac{1}{m q D}
	\left(
	\partial_1F\,\partial_2G
	-
	\partial_2F\,\partial_1G
	\right).
\end{equation}
Equivalently, on the critical surface \(m=-qs\),
\begin{equation}
	\label{eq:appC41}
	\{F,G\}_{\rm red}
	=
	\frac{1}{q^2sD}
	\left(
	\partial_1F\,\partial_2G
	-
	\partial_2F\,\partial_1G
	\right).
\end{equation}
In particular,  
\begin{equation}
	\label{eq:appC42}
	\{x^1,x^2\}_{\rm red}
	=
	-\frac{1}{m q D}
	=
	\frac{1}{q^2sD}.
\end{equation}

Since the reduced symplectic potential \(a_i(\tau,x)\dd x^i\) depends explicitly on
\(\tau\), the reduced Hamilton equations take the form
\begin{equation}
	\label{eq:appC43}
	\Omega_{ij}\dot x^j
	=
	\partial_i H_{\rm red}
	+
	\partial_\tau a_i .
\end{equation}
Equivalently,
\begin{equation}
	\label{eq:appC44}
	\dot x^i
	=
	\{x^i,H_{\rm red}\}_{\rm red}
	+
	\Pi^{ij}\partial_\tau a_j ,
\end{equation}
where \(\Pi^{ij}\) is the inverse Poisson tensor associated with
\(\Omega_{ij}\). Explicitly, away from $D=0$,
\begin{equation}
	\label{eq:appC45}
	\Pi^{12}
	=
	\{x^1,x^2\}_{\rm red}
	=
	-\frac{1}{m q D}.
\end{equation}

A similar procedure can be carried out for the massless actions.

\section{A useful identity}
\label{sec:useful-identity}

Let $(V,\eta, \vol)$ be an oriented lorentzian vector space and let $n
\in V$ be a null vector and $m \in V$ a unit-norm (spatial) vector
orthogonal to $n$.  It follows that $m,n$ are linearly independent.
The orientation and the inner product define a Hodge star operator
$\star : V \to \wedge^2 V$ defined by
\begin{equation}
	v \wedge \star w = \eta(v,w) \vol
\end{equation}
for all $v,w \in V$.   Since $m,n$ span the orthogonal complement
$n^\perp$, it follows that
\begin{equation}
	\star n = \lambda m \wedge n,
\end{equation}
for some $\lambda \in \RR$ which we can calculate as follows.

We introduce a vector $\ell \in V$ with $\eta(\ell, n) = 1$.  This
vector is uniquely defined modulo $n^\perp$.  Therefore on the one
hand
\begin{equation}
	\ell \wedge \star n = \eta(\ell, n) \vol = \vol
\end{equation}
and on the other hand this is equal to
\begin{equation}
	\lambda \ell \wedge m \wedge n.
\end{equation}
Therefore we see that
\begin{equation}
  \lambda \ell \wedge m \wedge n = \vol.
\end{equation}

Let $e_A = (e_0,e_1,e_2)$ be a basis for $V$ with $\eta_{AB}$ diagonal
with $\eta_{00} = -1$, $\eta_{11}  = \eta_{22}= 1$ and $\vol = - e_0
\wedge e_1 \wedge e_2$, so that $\vol = \tfrac16 \epsilon^{ABC} e_A
\wedge e_B \wedge e_C$, with $\epsilon^{012} = -1$.  Equivariance
under the proper Lorentz group $\SO(V) \cong \SO(2,1)$, which is the
subgroup of $\GL(V)$ preserving both the inner product $\eta$ and the
orientation $\vol$, allows us to choose without loss of generality $n
= e_0-e_1$, $m = e_2$ and $\ell = -e_1$, so that
\begin{equation}
  \lambda \ell \wedge m \wedge n = \lambda (- e_1) \wedge e_2 \wedge
  (e_0 - e_1) = - \lambda e_0 \wedge e_1 \wedge e_2 = \lambda \vol,
\end{equation}
which implies that $\lambda = 1$.  In other words,
\begin{equation}
  \star n = m \wedge n.
\end{equation}
We may now rewrite this identity relative to the basis $e_A$.  The LHS
is given by
\begin{equation}
  \star n = \tfrac12 \epsilon^{ABC} n_C e_A \wedge e_B
\end{equation}
as can be checked by ensuring that $\ell \wedge \star n = \vol$.  The
RHS is of course simply
\begin{equation}
  m \wedge n = \tfrac12 (m^A n^B - m^B n^A) e_A \wedge e_B.
\end{equation}
Therefore the identity becomes
\begin{equation}
  m^A n^B - m^B n^A = \epsilon^{ABC} n_C \iff m_A n_B - m_B n_A = \epsilon_{ABC} n^C.
\end{equation}

\end{appendices}

\bibliographystyle{JHEP}
\bibliography{MB_ref}

\end{document}